\documentclass[preprint,floatfix,epsfig]{revtex4}

\usepackage{graphicx} 
\usepackage{dcolumn} 
\usepackage{amsmath}
\usepackage{amssymb}
\usepackage{bm}
\usepackage[dvips]{color}
\usepackage{times}
\usepackage{ulem}
\usepackage[justification=raggedright]{caption}

\usepackage{subcaption}
\usepackage{tabularx}
\usepackage{appendix}
\usepackage{amstext}
\usepackage{array}

\usepackage{bbm}

\definecolor{wine-stain}{rgb}{0.5,0,0} 
\definecolor{bblue}{rgb}{0,0.0,0.5} 

\usepackage[colorlinks=true
,urlcolor=blue
,anchorcolor=blue
,citecolor=blue %red
,filecolor=blue
,linkcolor=wine-stain
,menucolor=blue
,pagecolor=blue
,linktocpage=true
,pdfproducer=medialab
,linktoc=all %section
]{hyperref}

\usepackage{booktabs}  % for \toprule, \midrule, and \bottomrule macros

\allowdisplaybreaks

\newcommand{\ncmd}{\newcommand}

\ncmd{\lt}{\left}
\ncmd{\rt}{\right}
\ncmd{\tr}[1]{\mbox{Tr}\lt({#1}\rt)}
\ncmd{\half}{\frac{1}{2}}
\ncmd{\eps}{\epsilon}
\ncmd{\veps}{\varepsilon}
\ncmd{\dgr}{\dagger}
\ncmd{\sig}{\sigma}
\ncmd{\gam}{\gamma}
\ncmd{\rtarw}{\rightarrow}
\ncmd{\Rt}{\Rightarrow}
\ncmd{\abs}[1]{\lt\cb{#1}\rt\cb}
\ncmd{\avg}[1]{\lt\lb{#1}\rt\rb}
\ncmd{\dl}{\delta}
\ncmd{\Dl}{\delta}
\ncmd{\sgn}[1]{\mbox{sgn}\lt(#1\rt)}
\ncmd{\kap}{\kappa}
\ncmd{\wtil}[1]{\widetilde{#1}}
\ncmd{\thrfr}{\therefore}
\ncmd{\eq}[1]{Eq. \eqref{#1}}
\ncmd{\fig}[1]{Fig. \ref{#1}}
\ncmd{\Lam}{\Lambda}
\ncmd{\lam}{\lambda}
\ncmd{\dow}{\partial}
\ncmd{\ordr}[1]{\mathcal{O}\lt(#1\rt)}
\ncmd{\dsty}{\displaystyle}
\ncmd{\alert}[1]{\color{red}{#1}}
\ncmd{\mc}{\mathcal}
\ncmd{\mbf}[1]{\mathbf{#1}}
\ncmd{\Deriv}[2]{\frac{d{#1}}{d{#2}}}
\ncmd{\ParDeriv}[2]{\frac{\partial{#1}}{\partial{#2}}}
\ncmd{\step}[1]{\Theta\lt(#1\rt)}
\ncmd{\td}{\tilde} 
\ncmd{\what}{\widehat}
\ncmd{\om}{\omega}
\ncmd{\Om}{\Omega}
\ncmd{\vrho}{\varrho}
\ncmd{\vsig}{\varsigma}
\ncmd{\vkap}{\varkappa}
\ncmd{\bqa}{\begin{eqnarray}} 
\ncmd{\eqa}{\end{eqnarray}}
\ncmd{\nn}{\nonumber \\}
\ncmd{\nnum}{\nonumber}
\ncmd{\comment}[1]{{\color{red}{#1}}}
\definecolor{new_color}{RGB}{50,155,0}

\renewcommand{\v}[1]{\ensuremath{\boldsymbol{#1}}} % for vectors

\ncmd{\lb}{\big<}
\ncmd{\rb}{\big>}
\ncmd{\cb}{\big|}

\makeatletter
\newcommand*{\rom}[1]{\expandafter\@slowromancap\romannumeral #1@}
\makeatother

\begin{document}

\title{
Horizon as Critical Phenomenon
}

\author{Sung-Sik Lee\\
\vspace{0.3cm}
{\normalsize{Department of Physics $\&$ Astronomy, 
McMaster University,}}\\
{\normalsize{1280 Main St. W., Hamilton ON L8S 4M1, Canada}}
\vspace{0.2cm}\\
{\normalsize{Perimeter Institute for Theoretical 
Physics,}}\\
{\normalsize{31 Caroline St. N., Waterloo ON N2L 2Y5, 
Canada}}
}

\date{\today}

\begin{abstract}

We show that renormalization group flow 
can be viewed as a gradual wave function collapse, 
where a quantum state associated with the action of field theory
evolves toward a final state that describes an IR fixed point.
The process of collapse is described by the radial evolution in the dual holographic theory.
If the theory is in the same phase as the assumed IR fixed point, 
the initial state is smoothly projected to the final state.
If in a different phase,
the initial state undergoes a phase transition 
which in turn gives rise to a horizon in the bulk geometry.
We demonstrate the connection between critical behavior and horizon in an example,
by deriving the bulk metrics that emerge in various phases of the $U(N)$ vector model in the large $N$ limit 
based on the holographic dual constructed from quantum renormalization group.
The gapped phase exhibits a geometry that smoothly ends at a finite proper distance in the radial direction.
The geometric distance in the radial direction measures a complexity :
the depth of renormalization group transformation
that is needed to project the generally entangled UV state to a direct product state in the IR.   
For gapless states, entanglement persistently spreads out to larger length scales,
and the initial state can not be projected to the direct product state.
The obstruction to smooth projection at charge neutral point
manifests itself as the long throat in the anti-de Sitter space.
The Poincare horizon at infinity marks the critical point
which exhibits a divergent length scale in the spread of entanglement.
For the gapless states with non-zero chemical potential, 
the bulk space  becomes the Lifshitz geometry
with the dynamical critical exponent two.
The identification of horizon as critical point may provide
an explanation for the universality of horizon.
We also discuss the structure of the bulk tensor network
that emerges from the quantum renormalization group.

\end{abstract}

\maketitle

%\newpage
%\tableofcontents

\newpage
%\twocolumngrid
{
\hypersetup{
    colorlinks,
    citecolor=black,
    filecolor=black,
    linkcolor=black,
    urlcolor=black
}
\tableofcontents
}
%\onecolumngrid

%\begin{thebibliography}{99}
%\end{thebibliography}

\section{Introduction}

The AdS/CFT correspondence\cite{Maldacena:1997re,Witten:1998qj,Gubser:1998bc} 
opened the door to address questions on gravity
from the perspectives of quantum field theory.
However, it is often difficult to make concrete progress in this direction 
because one usually does not have a full control on both sides of the duality.
In order to take full advantage of the duality, 
it is desired to have a first principle construction of holographic duals 
from field theories, which may allow one
to pose certain questions on  gravity in solvable field theories. 

The general link between boundary field theories
and bulk gravitational theories
is made through renormalization group (RG),
where the bulk variables are scale dependent couplings 
and the radial direction corresponds to a length scale\cite{1998PhLB..442..152A,deBoer:1999xf,Skenderis:2002wp}.
However, the RG suggested by holography is different from the Wilsonian RG 
in that the bulk equation of motion for the single-trace couplings 
captures the flow of multi-trace operators as well\cite{Heemskerk:2010hk,2011JHEP...08..051F}.
Furthermore, the bulk variables have intrinsic quantum fluctuations 
while the Wilsonian RG is deterministic. 
This hints that one has to promote the Wilsonian RG to quantum RG 
to make a precise connection with holography.
Quantum RG provides a prescription
to construct holographic duals for general field theories\cite{Lee2012,Lee:2012xba,Lee:2013dln}.
In quantum renormalization group,
RG flow is described as a dynamical system,
where couplings of field theories are promoted to quantum operators.
While only a subset of operators are kept,
quantum fluctuations of their couplings
capture other operators that are not 
explicitly included in the RG flow.
The fluctuations in RG path make the bulk theory quantum mechanical.
The bulk theory that governs the dynamical RG flow
includes gravity because the metric that sources
the energy-momentum tensor is also promoted to
dynamical variable\cite{Lee:2012xba,Lee:2013dln}.

The goal of this paper is to provide some insight 
into horizon by applying quantum RG 
to a solvable field theory.
We will show that horizons can be understood as critical phenomena,
where scale dependent couplings exhibit a non-analyticity
as a critical RG scale is approached.
A divergent length scale associated with spread of entanglement 
at the critical point gives rise to 
the divergent red shift at the horizon.
We demonstrate this by deriving the bulk geometry for a solvable field theory 
from quantum RG.
For earlier ideas on possible connections between black hole horizons and phase transitions,
see \cite{2003IJMPA..18.3587C} and references there-in.

%%In general, there is no local symmetry breaking order parameter for the critical phenomenon.
%%Instead, the locality itself is an order parameter. 

Here is an outline and the main results of the paper.
In section \ref{State}, 
we explain how quantum states
can be defined from actions
of field theories. 
This is different from the usual connection between
a $D$-dimensional action and a $(D-1)$-dimensional quantum state. 
Here, an action itself is promoted to a wave function,
and a $D$-dimensional action defines a $D$-dimensional quantum state. 
Then the partition function is written as an overlap between two quantum states,
called IR and UV states.
If the IR state is chosen to be a fixed point, 
the UV state is constructed from deformations turned on in a given field theory.
As deformations in the Wilsonian effective action are renormalized under the conventional RG flow,
the UV quantum state evolves under quantum RG flow.
The full theory with the deformations may or may not flow to the IR fixed point 
depending on the kind and strength of the deformations.   

In section \ref{collapse}, 
we explain that RG flow can be viewed as a wave function collapse 
where the UV state is gradually projected toward the IR state.
In section \ref{QRG},
we show that the wave function collapse is 
described by the radial evolution of a holographic dual via quantum RG,
and that the quantum Hamiltonian that generates the RG flow 
can be faithfully represented in basis states
constructed from actions which include only single-trace operators.
In doing so, we also explain the structure of tensor network in the bulk. 
The tensor network generated from quantum RG has no pre-imposed kinematic locality. 
Instead, the bulk space is made of a network of multi-local tensors of all sizes.
Dynamical locality in bulk is determined by action principle
which controls the strength of non-local tensors.

In section \ref{VM}, we apply quantum RG to the U(N) vector model,
which was previously studied numerically \cite{Lunts2015}.
In this paper, we obtain the analytic solution for the bulk equations of motion in the large $N$ limit.
The exact solution allows one to derive the bulk metric from the first principle construction. 
For this, a real space coarse graining is employed.
The IR fixed point is chosen to be a direct product state in the insulating phase,
and the kinetic (hopping) terms are regarded as deformation to the fixed point.  
The Hamiltonian that generates coarse graining
gradually removes entanglement in the UV state. 
When the theory with the deformation is still in the insulating phase, 
the UV state is smoothly projected to the direct product state. 
In this case, the bulk geometry is capped off at a finite proper distance in the radial direction.
The proper distance in the radial direction measures the depth 
of RG transformations that is necessary to 
remove entanglement entirely in the UV state.
This confirms the conjecture\cite{2014arXiv1402.5674S} 
that relates the radial distance in the bulk and complexity.
When the deformation is large 
that the system is no longer in the insulating phase,
the UV state can not be smoothly projected to the direct product state 
as entanglement in the UV state keeps spreading out to larger length scales under the RG flow.
The persistent entanglement gives rise to the extended geometry with an infinite proper distance in the radial direction.
The obstruction to smooth projection culminates in a non-analyticity of the scale dependent state
caused by the divergent length scale in the spread of entanglement. 
This can be viewed as dynamical critical behavior\cite{PhysRevLett.110.135704,PhysRevLett.115.140602}
as the quantum RG flow for Euclidean field theories is interpreted as 
a Hamiltonian flow under an imaginary time evolution.
The critical point gives rise to the Poincare horizon 
in the anti-de Sitter (AdS) space
or the Lifshitz geometry in the bulk,
depending on whether the chemical potential is turned on or off.
In sections \ref{DPT} and \ref{summary}, 
we conclude with summary and 
comments on universality of horizon and possible way to extend to Lorentzian space.

%%%%%%%%%%
%%%%NEW%%%
%%%%%%%%%%
Before we move on to the main body of the paper, 
we provide an intuitive picture of quantum RG 
with emphasis on the differences from
other approaches. 

\begin{itemize}
\item Wilsonian RG and Quantum RG :

An action is specified by the sources (couplings) for operators allowed within a theory.
The conventional Wilsonian RG describes 
the evolution of the sources under a coarse graining.
The Wilsonian RG flow is classical because
the sources at low energy scales
are determined without any uncertainty 
from the sources specified at high energy scales.
Beta functions, which are first order differential equations, govern the flow of the classical sources.
If the complete set of sources are kept within the flow,
the Wilsonian RG is exact.
However, it is usually difficult to keep all sources,
especially for strongly coupled theories
where there is no obvious way of truncating operators 
without knowing the spectrum of scaling dimensions beforehand.

Quantum RG is an alternative exact scheme which keeps only a 
subset of operators (single-trace operators).
It is based on the observation that a theory with arbitrary operators
can be represented as a linear superposition of theories 
which include only single-trace operators.
In other words, one can keep the exact information about an action with arbitrary sources
in terms of a quantum wave function defined in the space of the single-trace sources.
For example,  
the Boltzmann weight of a theory with operators, $O$, $O^2$, $O^3$, ...
can be represented as a linear superposition 
of the theories which have only $O$ through  
$e^{J_1 O + J_2 O^2 + J_3 O^3 + ... }  = \int dj_1 ~ \Psi(j_1) e^{j_1 O}$,
where the wave function $\Psi(j_1)$ is chosen
such that the original action which is non-linear in $O$ is reproduced 
upon integrating over $j_1$.
This reduces to the standard Hubbard-Stratonovich transformation
when the cubic and higher order terms are absent.
Then the exact RG flow generated from a coarse graining 
naturally becomes a quantum evolution of the wave function
because an action at any scale can be represented as a wave function.
The information on the infinite number of classical sources
is kept by the wave function of one source,
by taking advantage of the linear superposition principle.
In this sense, what quantum RG is to the conventional RG
is what quantum computer is to the classical computer.

The advantage of quantum RG is that one can keep 
only single-trace operators, 
which is typically a much smaller set 
compared with the full set of operators.
However, quantum RG is not necessarily simpler than the conventional RG for general theories
because one has to include quantum fluctuations of the single-trace sources.
The simplification arises when quantum fluctuations in RG paths are small
in the presence of a large number of degrees of freedom.
For this reason, quantum RG can be most useful 
in studying strongly interacting theories 
with large numbers of degrees of freedom.

\item Tensor networks and Quantum RG :

In quantum RG, the partition function is expressed
as a path integration of dynamical sources defined  
in one higher dimensional space,
where the extra dimension corresponds to the RG scale.
The path integral represents the sum over RG paths
defined in the space of single-trace sources.
The resulting path integral can be identified as a tensor network
where the bulk action becomes the constituent tensors
that need to be contracted through path  integral.
There are some characteristics 
of the tensor network generated from quantum RG,
which distinguish it from other forms of tensor networks.
First, the bulk tensor network that arises in quantum RG has
no pre-imposed locality.
This is due to the fact that the set of single-trace operators
in general include non-local operators.
The examples include the Wilson-loop operators in gauge theories
and the bi-local operators in vector models.
For this reason, the bulk space described by quantum RG is 
not only quantum but also non-local in general.
A classical and local geometry emerges only in a class of systems,
where quantum fluctuations of the dynamical sources are small,
and the sources for non-local operators decay fast enough at the saddle point.
Therefore, the degree of locality in the bulk is a dynamical feature
rather than a choice of local tensor structure put in by hand. 
Second, quantum RG chooses 
the dynamical sources and their conjugate momenta
as internal variables that are contracted (integrated over) in the bulk.
Because the source for the energy-momentum is the metric, 
the quantum RG naturally gives a bulk theory 
which includes dynamical gravity.
Although one may define a notion of distance for any local tensor network,
it generally does not represents a dynamical space 
if the connectivity of the network is rigid.

\end{itemize}

\section{Wave function of paths }
\label{State}

We consider a partition function in $D$-dimensional Euclidean space,
\bqa
Z = \int D\phi ~ e^{ - S },
\eqa
where  $\phi$ represents the fundamental field of the theory
and $S$ is the action.
If the $D$-dimensional partition function originates from 
a $(D-1)$-dimensional quantum theory,
the path integral sums over spacetime trajectories
in the Euclidean time.
On the other hand, the Boltzmann weight itself 
can be viewed as a $D$-dimensional wave function.
From an action $S$, one can define a quantum state 
\bqa
\cb S \rb &=& \int D \phi ~ e^{ -  S[\phi] } \cb\phi \rb, 
\label{action_state}
\eqa  
where $\cb \phi \rb$ is a complete set of basis states
which form a Hilbert space with the norm,
\bqa
\lb \phi^{'} \cb \phi \rb = \prod_{i,a} \delta( \phi^{'}_{ia} - \phi_{ia} ).
\eqa
Here $i$ is an index for the $D$-dimensional space,
and $a$ represents internal flavour(s).
For concreteness, one can use a lattice regularization for the $D$-dimensional space. 
From the point of view of the original $(D-1)$-dimensional quantum theory,
\eq{action_state} represents a state of spacetime path.

Any quantity that can be defined for usual quantum states
can be defined for the states generated from actions.
For example, the von Neumann entanglement entropy 
can be defined for the $D$-dimensional states
as it is defined for the $(D-1)$-dimensional state 
for the original quantum system.
However, the physical meanings of the two objects are somewhat different.
The $D$-dimensional state
encodes entanglement between (Euclidean) spacetime {\it events}
rather than entanglement between the {\it bits} that reside in the $(D-1)$-dimensional space.

\begin{figure}[h]
\centering
\includegraphics[width = 0.5\columnwidth]{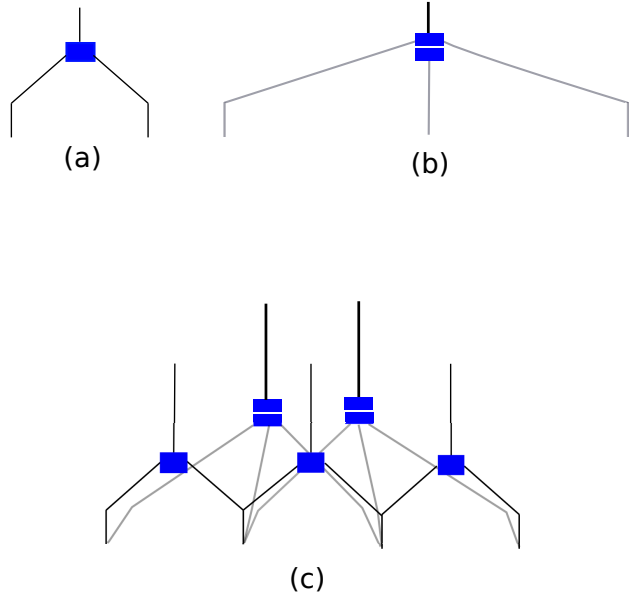}
\caption{
(a) An example of tensor constructed from an operator $O_{ij}$ that depends on fields $\phi_i$, $\phi_j$. 
The solid box represents a rank $3$ tensor, $T^{J}_{\phi_i, \phi_j} = e^{J O_{ij} }$,
where the leg above the box is $J$
and the two legs below the box are $\phi_i$, $\phi_j$. 
(b) An example of tensor made of a composite operator,
$T^{J}_{\phi_i, \phi_j, \phi_k} = e^{ J O_{ij} O_{jk} }$,
which has support on three points in space.
The four legs correspond to $J$, $\phi_i$, $\phi_j$ and $\phi_k$.
(c) A quantum state associated with an action can be written as a direct product of tensors,
where the sources in the action become variational parameters of the state.
}
\label{fig:1}
\end{figure}

An action can be written as
\bqa
S = -\sum_i J_i O_i - \sum_{i,j} J_{ij} O_{i,j} + ...,
\eqa
where $O_{i_1,i_2,..,i_n}$'s are operators  
that depends on $n$ space coordinates,
and  $J_{i_1,i_2,..,i_n}$'s are the sources.
For notational simplicity, we suppress the indices for internal flavours.
One can identify the Boltzmann weight associated with a $n$-point operator 
as a rank $(n+1)$ tensor,
\bqa 
T^{J}_{\phi_{i_1},\phi_{i_2},..,\phi_{i_n}}
= 
e^{  J O_{i_1,i_2,..,i_n}   }.
\label{TJ}
\eqa
Here one of the indices in the tensor
is the source, $J$,
and the remaining indices are the values of the fields
at $n$ points, $\phi_{i_1}$, $\phi_{i_2}$,..,$\phi_{i_n}$. 
The indices are in general  continuous.
The $D$-dimensional state is then written as 
\bqa
\cb S \rb &=& \int D \phi ~ \Psi_J[\phi] \cb\phi \rb,
\eqa
where the wave function is given by a direct product of the multi-local tensors,
\bqa
\Psi_{J}[\phi] & = & \prod_n  \prod_{i_1,..,i_n}  T^{J_{i_1,i_2,..,i_n}}_{\phi_{i_1},\phi_{i_2},..,\phi_{i_n}}.
\eqa
It is noted that $\Psi_J[\phi]$ is a wave function of $\phi$.
The sources are variational parameters in the wave function.
These are illustrated in \fig{fig:1}.
For local actions, the sources for multi-local operators decay exponentially 
as the separation between coordinates increases.
States generated from local actions are called {\it local states}.

If there is a symmetry in the action,
$S[\phi]$ is invariant under the symmetry transformations,
\bqa
S[ U_s \phi] = S[\phi].
\eqa
Here $U_s$ is a matrix that acts on $\phi$
which is viewed as a giant vector,
where space points and internal indices form the vector index.
The corresponding quantum state $\cb S \rb$  
is  invariant under a unitary transformation $\hat U_s$
which acts on the basis states as
\bqa
\hat U_s \cb \phi \rb = \cb U_s \phi \rb.
\eqa
We call states generated from symmetric actions {\it symmetric states}.
The set of symmetric states form {\it symmetric Hilbert space}.

A symmetric action can be written as a sum of singlet operators with sources, 
\bqa
S & = &  - {\cal J}^M {\cal O}_M.
\eqa
Here $\{ {\cal O}_M \}$ represents a complete set of singlet operators
that are invariant under the symmetry of the theory. 
Here $M$, which is summed over, denotes not only different types of operator, 
but also operators at different positions in $D$ dimensions.
In general, ${\cal O}_M$ can be multi-local operators that depend on multiple coordinates.
In matrix models, for example, a trace of matrix fields at multiple positions
and product of them form singlet operators. 
${\cal J}^M$ are the sources.
Now symmetric states can be labeled
by the sources of the singlet operators,
\bqa
\cb \{ {\cal J} \}   \rb &=& \int D \phi ~ e^{  {\cal J}^M {\cal O}_M  } \cb\phi \rb.
\label{action_state2}
\eqa  
However, these states form an over-complete basis of the symmetric Hilbert space.
There is a smaller set of complete basis.
The minimal set of operators that span the full symmetric Hilbert space
are called single-trace operators, $\{ O_n \}$.
In large $N$ matrix models, they are the operators that involve one trace.
But the definition of single-trace operator can be extended to any theory.
In general, $\{ O_n \}$ is defined to be the minimal set of singlet operators 
of which all singlet operators can be expressed as  polynomials\cite{Lee:2013dln},
\bqa
{\cal O}_M = \sum c^{n_1, n_2,..}_M  O_{n_1} O_{n_2}..
\eqa

\begin{figure}[h]
\centering
\includegraphics[width = 0.3\columnwidth]{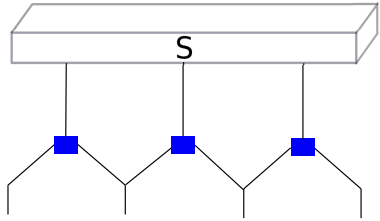}
\caption{
%A symmetric state can be written as a linear superposition of the basis states
%constructed from single-trace operators without using the tensors from multi-trace operators.
The state in \fig{fig:1}(c) represented in the single-trace basis.
The solid boxes represent tensors constructed from the single-trace operators.
The $3D$ box represents the wave function written in the basis of the single-trace tensors.
The legs associated with the sources are contracted.  
}
\label{fig:general_state}
\end{figure}

The single-trace operators span the full symmetric Hilbert space because 
\bqa
\cb S \rb & = & \int D \phi ~ e^{ \sum_{k} {\cal J}^{n_1,n_2,..,n_k} O_{n_1} O_{n_2} .. O_{n_k}        } ~ \cb\phi\rb  \nn
& = & \int Dj D\phi ~ e^{ -j^n ( j_n^* - O_n ) + \sum_{k} {\cal J}^{n_1,n_2,..,n_k} j_{n_1}^* j_{n_2}^* .. j_{n_k}^*        
} ~ \cb\phi\rb.
\label{eq:Sp}
\eqa
Here $D j \equiv \prod_n dj^n dj_n^*$. 
$j^n, j_n^*$ are auxiliary fields introduced for each single-trace operator,
where $j_n^*$ is the complex conjugate of $j^n$.
We used the identity $f(O)=\frac{1}{\pi}\int dj~dj^*~e^{-j(j^*-O)}~f( j^*)$\cite{Lee2012}.
A multiplicative constant that is independent of sources is ignored in \eq{eq:Sp}.
\eq{eq:Sp} shows that any singlet state 
can be written as a linear superposition of the states 
constructed from the single-trace operators as
\bqa
\cb S \rb & = & \int Dj ~ \Psi_S( j)~ \cb j \rb,
\eqa
where
\bqa
\cb j \rb & = & \int D \phi ~e^{  j^n O_n} \cb \phi \rb
\eqa
is the complete single-trace basis and 
\bqa
\Psi_S(j) &=&  e^{  - j_n^* j^n  + \sum_{k} {\cal J}^{n_1,n_2,..,n_k} j_{n_1}^* j_{n_2}^* .. j_{n_k}^*        }
\eqa
is the wave function written in the space of $j$.
Graphically, symmetric states are made of
single-trace tensors whose sources
are contracted with the tensor given by $\Psi_S(j)$
(\fig{fig:general_state}).

Therefore it is natural to associate an action of a field theory
with a wave function.
Instead of specifying classical sources for all singlet operators,
one can specify a quantum wave function defined in the space of 
the single-trace sources.
The quantum RG is based on the fact that
the exact Wilsonian RG flow defined in the classical space of full singlet operators
can be faithfully represented as a quantum evolution
of wave function defined in the space of single-trace operators\cite{Lee:2013dln}.
In the following section, we will first show that RG flow can be 
viewed as a collapse of such wave function defined from an action.

\section{Coarse graining as wave function collapse }
\label{collapse}

%%\subsection{Partition function as an overlap of wave functions}

\begin{figure}[h]
\centering
\includegraphics[width = 0.3\columnwidth]{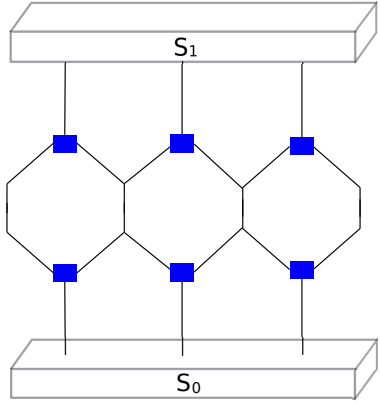}
\caption{
The partition function can be viewed as the overlap between two quantum states.
The lines contracted with $S_0$ and $S_1$ are the sources,
and the lines that are contracted between solid boxes (single-trace tensors)
are the fundamental fields of the theory.  
}
\label{fig:Z}
\end{figure}

The goal of renormalization group 
is to integrate out degrees of freedom partially,
and examine how the action for the remaining degrees of freedom flow.
The order in which the degrees of freedom are integrated out 
is determined by a reference action $S_0$.
It is convenient to choose $S_0$ to be a fixed point of the theory 
in order to keep the reference action invariant under the scale transformation. 
If $S_0$ is chosen to be the free kinetic term as is usually done in perturbative RG,
the order of integration is organized according to momentum.
However, there is freedom to choose different reference action.
Once the choice is made for the reference action,
the remaining action, $S_1 = S-S_0$ is regarded as 
a deformation to the reference action.

Now we define quantum states associated with the reference action and the deformation
in a given field theory as 
\bqa
\cb S_0^*   \rb &=& \int D \phi ~ e^{ -  S_0^*[\phi] } \cb\phi \rb, \nn
\cb S_1  \rb & = & \int D \phi ~ e^{ - S_1[\phi] } \cb\phi \rb.
\eqa  
$\cb S_0^* \rb$ and $\cb S_1 \rb$ are called 
IR and UV states respectively.
The reason why we define the IR state from  the complex conjugate of the reference action 
is because the partition function for the full action $S=S_0+S_1$ 
is  given by the overlap between the two states,
\bqa
Z = \lb S_0^* \cb S_1 \rb.
\eqa
If $\cb S_0^* \rb$ and $\cb S_1\rb$ are written in the single-trace basis, 
$\cb S_0^*  \rb  =  \int D j^{'} ~ \Psi_{0}(j^{'})  \cb j^{'} \rb$,
$\cb S_1  \rb  =  \int D j ~ \Psi_{1}(j)  \cb j \rb$,
the partition function becomes 
\bqa
Z & = & \int Dj^{'} Dj ~  \Psi^*_{0}(j^{'})   \lb j^{'} \cb j \rb \Psi_{1}( j ).
\eqa
This is illustrated in \fig{fig:Z}.

A coarse graining procedure in the renormalization group transformation
can be viewed as a collapse of the UV state towards the IR state.
To illustrate the idea, let us begin with a simple example of one real scalar integration,
\bqa
Z = \int d \phi ~ e^{-S_0 - S_1}.
\eqa
Here $S_0 = \frac{m^2}{2}  \phi^2$ is the quadratic reference action.
All the higher order terms are included in the deformation $S_1[\phi]$.
In the presence of the $Z_2$ symmetry, $\phi \rightarrow -\phi$,
only even powers are allowed in the action.
For coarse graining, one adds an auxiliary field $\Phi$ with a mass $\mu$.
The physical field and the auxiliary field are mixed into a new basis,
$\phi = \phi^{'} + \Phi^{'}$,
$\Phi = \frac{m}{\mu \sqrt{2 dz} } ( -2 dz \phi^{'} + \Phi^{'} )$,
in which the new action becomes
\bqa
S = \frac{m^2 e^{2 dz}}{2} \phi^{'2} + \frac{m^2}{4 dz} \Phi^{'2} + S_1[ \phi^{'} + \Phi^{'}].
\eqa  
The fluctuations of $\phi^{'}$ are slightly suppressed compared to the original field $\phi$
due to the slightly increased mass.
The missing fluctuations are carried by $\Phi^{'}$.
Therefore, integrating out $\Phi^{'}$ 
has the effect of partially including fluctuations of $\phi$.
It generates quantum corrections, 
\bqa
\delta_1 S_1 & = & \frac{dz }{m^2}  \left[
\frac{\partial^2 S_1}{\partial \phi^{'2} }
-
\left( \frac{\partial S_1}{\partial \phi^{'}} \right)^2
\right].
\eqa
Rescaling of the field, $\phi^{'} = e^{ -dz} \phi^{''}$,
brings $S_0$ back to the original form,
 and generates an additional correction from $S_1$,
\bqa
\delta_2 S_1 & = & -dz \phi^{''} ~\frac{\partial S_1}{\partial \phi^{''}}.   
\eqa
The relation between the original field $\phi$ and the low energy field $\phi^{''}$ is given by
\bqa
\phi^{''} = e^{- dz} \left( \phi - \frac{\mu}{m} \sqrt{2 dz} \Phi \right).
\label{OL} 
\eqa
Renaming $\phi^{''}$ to $\phi$,
we note that the net quantum correction to $S_1$ is generated by a quantum evolution,
\bqa
e^{-(S_1 + \delta_1 S_1 + \delta_2 S_1 )} =
\lb \phi \cb e^{-dz \hat H} \cb S_1 \rb,
\eqa
where
$\cb S_1 \rb = \int d \phi ~e^{-S_1[\phi] }  \cb \phi \rb$ is the quantum state
corresponding to the action $S_1$, and
\bqa
\hat H = 
 \frac{1 }{m^2} \hat \pi^2
+ i \hat \phi \hat \pi
\eqa
is the Hamiltonian.
$\hat \phi$ has eigenvalue $\phi$ for $\cb \phi \rb$, $\hat \phi \cb \phi \rb = \phi \cb \phi \rb$.
$\hat \pi$ is the conjugate momentum that obeys the commutation relation,
$[\hat \phi, \hat \pi ] = i$. 

Here $\hat H$ is the generator of the renormalization group transformation.
Although $\hat H$ is not a Hermitian Hamiltonian,
it is symmetric under the combined transformation of 
the parity (P), $\phi \rightarrow -\phi, \pi \rightarrow -\pi$
and the time reversal (T), $\phi \rightarrow \phi, \pi \rightarrow -\pi, i \rightarrow -i$.
The PT-symmetry guarantees that both $\hat H$ and $\hat H^\dagger$ have real eigenvalues\cite{2007RPPh...70..947B}.
However, $\hat H$ and $\hat H^\dagger$ have different eigenstates which are related to each other
through a similarity transformation.
The ground state wave function of $\hat  H^\dagger$ ($\hat H$) is $e^{-S_0}$ ($1$)
with the ground state energy $0$. 
Since $\hat H^\dagger$ annihilates $\cb S_0^* \rb = \cb S_0 \rb$,
the IR state is invariant under the transformation, 
which reflects the fact that $S_0$ is a fixed point.
Therefore, the partition function is invariant under the  insertion of $e^{-dz \hat H}$.
On the other hand, $e^{-dz \hat H}$ acting on $\cb S_1 \rb$ 
generates a non-trivial evolution of the deformation,
\bqa
Z  =  \lb S_0^* \cb S_1 \rb = \lb S_0^* \cb e^{-dz \hat H} \cb S_1 \rb =  \lb S_0^* \cb  S_1+ \delta S_1 \rb. 
\label{dzH}
\eqa
Now an infinite sequence of $e^{-dz \hat H}$ is inserted to write
\bqa
Z =  \lim_{z \rightarrow \infty} \lb S_0^* \cb e^{-z  \hat H} \cb S_1 \rb.
\eqa
Since  $\hat H$ has the unique ground state 
with a finite gap in the spectrum,
$\cb S_1 \rb$ is gradually projected toward the ground state in the long time limit,
where we interpret $z$ as an imaginary time. 
The process of projecting the state $\cb S_1 \rb$ to the ground state is the RG flow.
In the large $z$ limit, $e^{-z \hat H} \cb S_1 \rb$ approaches the ground state of $\hat H$,
which has a trivial overlap with $\cb S_0^* \rb$.
The information on the partition function is encoded in the norm of the projected state.

In the above example, the RG flow is smooth.
Any state with a finite number of degrees of freedom is smoothly projected to the ground state.
However, this is not true in the thermodynamic limit
because the large system size limit and the large RG time limit
do not commute in general.
As will be discussed in Sec. \ref{VM},
$\cb S_1 \rb$ can not be smoothly projected to the ground state 
in the thermodynamic limit
if the deformation is large that the system 
is no longer in the phase described by the assumed IR fixed point
described by $S_0$.

Here let us consider another simple example 
to relate RG flow with wave function collapse.
We consider a $D$-dimensional scalar field theory,
where the reference action is chosen to be the free kinetic term,
\bqa
S_0 = \frac{1}{2} \int d^D k~ G^{-1}_\Lambda(k) \phi_k \phi_{-k}.
\eqa
Here $k$ is momentum, and
$G^{-1}_\Lambda(k)$ is a regularized kinetic term with UV cut-off $\Lambda$, 
e.g., $G^{-1}_\Lambda(k) = e^{ \frac{k^2}{\Lambda^2} } k^2$.
All the interactions are included in the deformation, $S_1[\phi]$.
After lowering the cut-off $\Lambda \rightarrow \Lambda e^{-dz}$
followed by a rescaling of field and momentum,
$\phi_k \rightarrow e^{ \frac{D+2}{2} dz} \phi_{ e^{dz} k}$,
one obtains a quantum correction $\delta S_1$\cite{Polchinski:1983gv}.
The quantum correction can be generated by 
\bqa
e^{-(S_1[\phi] + \delta S_1[\phi] )} =
\lb \phi \cb e^{-dz \hat H} \cb S_1 \rb,
\eqa
where
\bqa
\hat H = 
 \int dk \left[
  \frac{\td G(k)}{2}  
\hat \pi_k \hat \pi_{-k}
-
i \left( \frac{D+2}{2}  \hat \phi_k + k \partial_k  \hat \phi_k  \right) \hat \pi_{-k}
\right]
\label{H2}
\eqa
with $\td G(k) = \frac{\partial G_\Lambda(k)}{\partial \ln \Lambda}$.
The conjugate momentum obeys the commutation relation,
$[\phi_k, \pi_{k'} ] = i \delta(k+k^{'})$. 
The first term in \eq{H2} is from lowering $\Lambda$,
and the second term originates from the rescaling of field and momentum.
However, the validity of \eq{dzH} is subtle in this case
because $\cb S_0 \rb$ is not exactly annihilated by $\hat H^\dagger$ in general.
Upon applying $\hat H^\dagger$ to $\cb S_0 \rb$, 
one is left with a total derivative term, 
$-\frac{1}{2} \int d^D  k ~ \nabla_k \cdot ( \vec k G^{-1} \phi_k \phi_{-k} )$.
This is due to the fact that the range of momentum changes 
as momentum is relabeled.
The total derivative term can be dropped if
the range of momentum is formally taken to infinity.
However, the partition function is not finite in this case.
To avoid this subtlety, we will adopt a real space renormalization group scheme in the following.
Another advantage of using a real space RG scheme is the easiness 
in implementing a local RG\cite{1991NuPhB.363..486O,Lee:2012xba,Lee:2013dln,2015arXiv150207049N}  
by choosing the speed of coarse graining 
differently at different points in space.

\section{Quantum renormalization group}
\label{QRG}

\begin{figure}[h]
\centering
\includegraphics[width = 0.3\columnwidth]{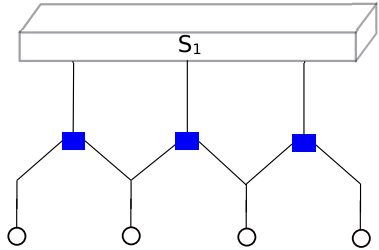}
\caption{
The same figure as in \fig{fig:Z} except that the reference state 
$\cb S_0^* \rb$ is chosen to be a direct product state in real space.
Each circle represents a rank one tensor which depends on  $\phi_i$ at each site.  
}
\label{fig:Z_local}
\end{figure}

In the real space renormalization group scheme, 
we can choose $\cb S_0^* \rb$ to be a direct product state.
Accordingly, the Hamiltonian 
of which $\cb S_0^* \rb$ is the ground state (with zero energy)
is ultra-local,
\bqa
\hat H = \sum_i \alpha_i \hat h_i,
\label{eq:H}
\eqa
where 
$\hat h_i$ is the on-site term of the Hamiltonian,
and $\alpha_i$ is a local speed of coarse graining.
We note that the choice of $\hat h_i$ is not unique for a given $\cb S_0 \rb$.
Different Hamiltonians project excited states in different rates,
which corresponds to use different RG scheme. 
Furthermore, one can choose any $\alpha_i > 0$
because $\cb S_0^* \rb$ is a direct product state.
%%This is a gauge freedom which is a part of the bulk diffeomorphism\cite{Lee:2012xba}.
Here we will choose a gauge $\alpha_i=1$ for simplicity. 
The partition function is given by
\bqa
Z = \int Dj^{(0)}~  \lb S_0^* \cb j^{(0)} \rb \Psi( j^{(0)} ),
\eqa 
where  $\Psi( j^{(0)} )$ is the wave function associated with the UV state
as is illustrated in \fig{fig:Z_local}.

\begin{figure}[h]
\centering
\includegraphics[width = 0.75\columnwidth]{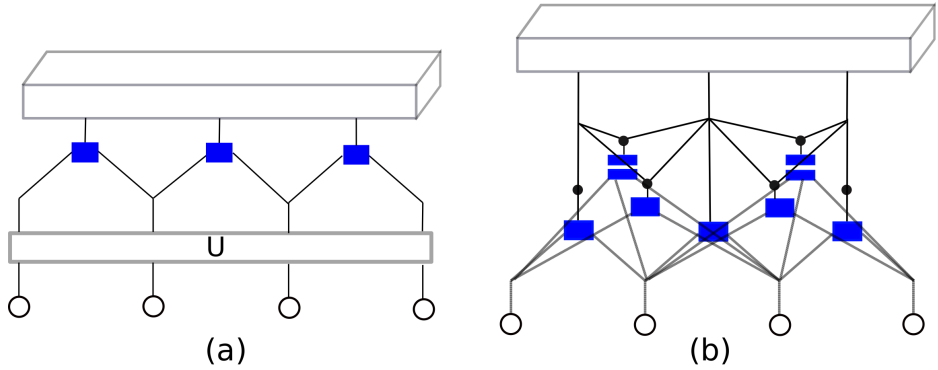}
\caption{
(a) One inserts $U = e^{- dz \hat H}$ between the overlap $\lb S_0^* \cb S_1 \rb$,
where $\hat H$ is chosen such that 
$\cb S_0^* \rb$ is the ground state of $\hat H^\dagger$ with zero eigenvalue.
(b) $U$ generates a non-trivial evolution of the UV state. 
$U \cb S_1 \rb$ includes
not only more non-local single-trace tensors (solid boxes) than what was present in $\cb S_1 \rb$,
but also multi-trace tensors (boxes with stripes).
The sources for these new tensors are determined by the original single-trace sources.
For example, the striped box represents a double-trace tensor of rank $4$,
$T^{{\cal J}_{ijk}}_{\phi_i, \phi_j, \phi_k} = e^{ {\cal J}_{ijk} O_{ij} O_{jk} }$,
and the solid dot above the striped box represents 
a rank $3$ tensor $\tilde T^{j^{(0)}_{ij},j^{(0)}_{jk}}_{{\cal J}_{ijk}} = \delta \left( {\cal J}_{ijk} - dz C^{ij, jk}[j^{(0)}] \right)$,   
which relates ${\cal J}_{ijk}$ with the sources of the original single-trace operators,
where $C^{ij, jk}[j^{(0)}]$ is a function of $j^{(0)}_{ij}, j^{(0)}_{jk}$.
}
\label{fig:E1}
\end{figure}

Since $\cb S_0^* \rb$ is annihilated by $\hat H^\dagger$, 
the overlap is invariant under the insertion of
$U= e^{-dz \hat H}$ (Fig. \ref{fig:E1} (a)),
\bqa
Z = \int Dj^{(0)}~ \lb S_0^* \cb e^{-dz \hat H } \cb j^{(0)} \rb \Psi( j^{(0)} ).
\eqa 
Even though $\cb j^{(0)} \rb$ includes only single-trace tensors,
the evolution generates multi-trace tensors,
\bqa
e^{-dz  \hat H} \cb j^{(0)} \rb = 
\int D \phi ~~
e^{ j^{(0) n} O_n - dz  C^{n_1,n_2,..}[j^{(0)}] \left( O_{n_1} O_{n_2}... \right) }
\cb\phi\rb. 
\eqa
Furthermore, longer-range tensors are generated from short-range tensors.
The sources for the longer-range tensors and multi-trace tensors
are functions of $j^{(0)}$ as is illustrated in Fig. \ref{fig:E1} (b).

\begin{figure}[h]
\centering
\includegraphics[width = 0.75\columnwidth]{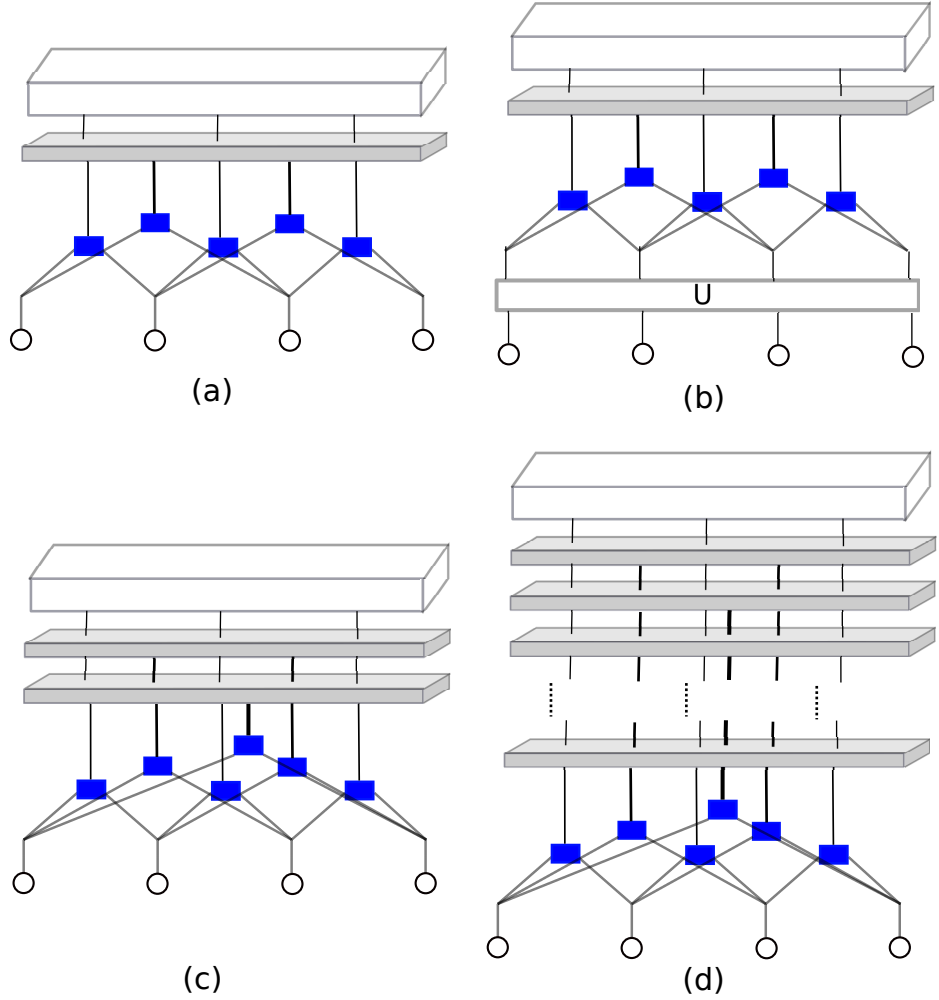}
\caption{
Bulk tensor network generated from quantum RG.
(a) $U \cb S_1 \rb$ in \fig{fig:E1}(b) can be written as a linear superposition
of states made of single-trace tensors.
Here the sources for the single-trace tensors become dynamics.
(b) $U$ is inserted again.
(c) The evolution from $U$ generates even more non-local single-trace 
tensors along with multi-trace tensors.
Multi-trace tensors are removed by promoting the
sources for the single-trace tensors dynamical.
(d) Repeating the steps (b) and (c), generates
a $(D+1)$-dimensional network of tensors,
where dynamical sources are contracted. 
In the bulk, non-local single-trace tensors of all sizes are generated
even if only local tensors are present at the UV boundary.
}
\label{fig:E2}
\end{figure}

Because general symmetric states can be represented as a linear combination of  $\cb j \rb$,
the Hamiltonian should have a faithful representation in the space of $j$. 
The state with multi-trace operators can be again written as a linear combination of the single-trace states,
\bqa
e^{-dz \hat H} \cb j^{(0)} \rb = 
\int  D j^{(1)}  ~~
e^{   - j^{*(1)}_n (  j^{(1)n} -  j^{(0) n} ) 
- dz C^{n_1,n_2,..}[j^{(0)}] \left( j^{*(1)}_{n_1} j^{*(1)}_{n_2}... \right)
 }
\cb j^{(1)} \rb. \nn
\eqa
In Fig. \ref{fig:E2}(a), the gray box represents the tensor,
$
e^{   - j^{*(1)}_n (  j^{(1)n} -  j^{(0) n} ) 
- dz C^{n_1,n_2,..}[j^{(0)}] \left( j^{*(1)}_{n_1} j^{*(1)}_{n_2}... \right)
 }
$, whose external legs
are $j^{(0)}$ and $j^{(1)}$.
By applying this successively,
the partition function can be written as a path integration of 
auxiliary sources $j^{(n)}$ introduced at $n$-th step of coarse graining.
This is illustrated in Fig. \ref{fig:E2} (b)-(d).
In the continuum ($dz \rightarrow 0$) limit, one obtains
\bqa
Z & =  & 
\lim_{z \rightarrow \infty}
\lb S_0^* \cb e^{ - z \hat H } \cb S_1 \rb, \nn
& = & 
\int Dj(z) ~~ \lb S_0^* \cb j(\infty)  \rb  ~ 
e^{ 
- \int_0^{\infty} dz  \left( 
 j_n^* \partial_z j^n +   
  {\cal H}[j^*,j] \right)
 } \Psi(j(0)).
 \label{eq:PI}
\eqa
${\cal H}[j^*,j]$ is the coherent state representation
of the quantum Hamiltonian,
\bqa
\hat {\cal H}[j^\dagger,j] & = &  
\left( j^{\dagger}_{n_1} j^{\dagger}_{n_2}... \right)
C^{n_1,n_2,..}[j].
\eqa
$j^n$, $j_n^\dagger$ are quantum operators
defined in the space of single-trace operators.
The space of single-trace operators depends on the type of field theory.
For gauge theories, it is the space of loops\cite{Lee2012}.
For vector models, it is the space of bi-local coordinates\cite{Lunts2015}.
The operators satisfy the commutation relation $[j^n, j_m^\dagger] = \delta^n_{m}$.
$j_n^\dagger$ ($j^n$) creates (annihilates) a closed string in gauge theories,
and a bi-local object in vector models. 
Because the complete set of single-trace operators include extended objects,
the bulk theory is kinematically non-local.
$\hat {\cal H}$ describes quantum evolution of those extended objects in the bulk.

The Hamiltonian gradually projects the generally entangled UV state to the direct product state.
The bulk theory can be viewed as a process of removing entanglement present in the UV state.
For other approaches of entanglement based renormalization group, see Refs. \cite{PhysRevLett.99.120601,2013PhRvL.110j0402H,2015PhRvL.115r0405E,2015PTEP.2015g3B03M}.
In the present case, the disentangler $e^{-z \hat H}$ is a non-unitary operator.
On the other hand, we can not choose any local operator to remove short-distance entanglement.
The Hamiltonian should annihilate the reference state in order to keep the overlap invariant.

Since $\hat H$ is an ultra-local Hamiltonian given by sum of on-site terms,
${\cal H}$ has an important property.
If one expands $C^{n_1,n_2,..}[j]$ in power series of $j$ as
\bqa
C^{n_1,n_2,..}[j] = \sum_{m_1,m_2,..}~  f^{n_1,n_2,..}_{m_1,m_2,..} ~j^{m_1} j^{m_2} ..,
\label{eq:C}
\eqa
the ultra-locality implies that $f^{n_1,n_2,..}_{m_1,m_2,..}$ is non-zero 
only if for every site $i$ included in the support of $\{ n_k \}$
there exists some $m_l$ whose support includes $i$.
In other words, $\hat {\cal H}$ can create an object $n$ by $j_n^\dagger$ 
only when there are pre-existing objects whose support include the support of $n$. 
This restriction can be understood as arising from a local symmetry that is present in the Hamiltonian\cite{Lee2012}.
We will illustrate this point further in the next section through a concrete example.

The path integral in \eq{eq:PI} is nothing but contractions
of tensors that form a network in the bulk 
as is illustrated in \fig{fig:E2}.
The internal variables contracted in the bulk are the dynamical sources. 
In the large $N$ limit, quantum fluctuations are suppressed, 
and the contractions of tensors are dominated by the saddle-point\cite{Lee:2012xba}.
The tensor network generated from quantum RG
has some differences from other forms of tensor networks\cite{PhysRevLett.101.110501,PhysRevD.86.065007,2012JHEP...10..193N,2013arXiv1309.6282Q,2013PhRvL.110j0402H,2015PhRvL.115q1602M, 2015JHEP...06..149P},
some of which have been proposed for holography.
The tensor network described in \eq{eq:PI} has no pre-imposed kinematic locality in the bulk
because one has to include tensors associated with multi-local single-trace operators of all sizes.
This kinematic non-locality is crucial in order to have a sense of 
diffeomorphism invariance in the bulk\cite{Lunts2015,2015PhRvL.114c1104M}.
In the absence of kinematic locality in the bulk,
the degree of locality in the classical geometry that emerges in the large $N$ limit
is determined dynamically.
In this sense, locality in the bulk is a dynamical feature of the theory
rather than a pre-imposed structure built in the tensor network.

\section{Examples : vector model}
\label{VM}

As a concrete application, 
we consider the $D$-dimensional vector model, 
\bqa
\mathcal{S} & = &  m^2\sum_i \left(\v{\phi}^*_i\cdot\v{\phi}_i\right)
 -\sum_{ij} {t}^{(0)}_{ij} \left(\v{\phi}^*_i\cdot\v{\phi}_j\right)
+ \frac{\lambda}{N}\sum_i\left(\v{\phi}^*_i\cdot\v{\phi}_i\right)^2.
\label{S1} 
\eqa
Here $\v\phi$ is complex boson field with $N$ flavours, 
$m$ is the mass of the bosons, and  $\lambda$ is the quartic coupling.
$t^{(0)}$ is the hopping parameter, which gives the kinetic term.
$i,j$ refer to site indices in a $D$-dimensional lattice.
They are promoted to continuous coordinates in continuum space. 
For other holographic approaches to the vector models
and related conjectures see Refs. 
\cite{2002PhLB..550..213K,
Das:2003vw,
Koch:2010cy,
Douglas:2010rc,
2013arXiv1303.6641P,
Leigh:2014tza,
2015PhRvD..91b6002L,
2014arXiv1411.3151M,
Vasiliev:1995dn,
Vasiliev:1999ba,
Giombi:2009wh,
Vasiliev:2003ev,
Maldacena:2011jn,
Maldacena:2012sf,
2014PhRvD..90h5003S}.

We write the partition function as
\bqa
Z = \lb S_0^* \cb t^{(0)} \rb,
\eqa
where 
\bqa
\cb S_0 \rb & = &  \int D \v{\phi} ~ e^{ -m^2\sum_i \v{\phi}^*_i\cdot\v{\phi}_i } \cb \v{\phi} \rb, \nn
\cb t^{(0)} \rb & = & \int D \v{\phi} ~ e^{ ~
 \sum_{ij} t^{(0)}_{ij} \v{\phi}_i^* \cdot \v{\phi}_j 
-\frac{\lambda}{N}\sum_i\left(\v{\phi}^*_i\cdot\v{\phi}_i\right)^2
} \cb \v{\phi} \rb.
%%, \nn \Psi( t^{(0)} ) & = &  e^{ N \sum_{ij} ( - t^{*(0)}_{ij} t^{(0)}_{ij} +  \td t_{ij} t^{*(0)}_{ij} ) }. 
\eqa
Because $S_0$ is real, $\cb S_0^* \rb = \cb S_0 \rb$.
In the vector model, the single-trace operators are the bi-local operators, 
$ \left(\v{\phi}^*_i\cdot\v{\phi}_j\right)$\cite{Das:2003vw,Koch:2010cy,2014arXiv1411.3151M}.
In $\cb t^{(0)} \rb$, one might try to remove the quartic double-trace operator 
by introducing an auxiliary field,
\bqa
\cb t^{(0)} \rb & = & \int  D \v{\phi} Dt   ~ e^{ ~
 \sum_{ij} t^{}_{ij} \v{\phi}_i^* \cdot \v{\phi}_j
 - N t^*_{ij}( t_{ij} - t^{(0)}_{ij} ) 
-N \lambda \sum_i (t_{ii}^*)^2
} \cb \v{\phi} \rb.
\eqa
However, this is ill-defined for finite $N$
where $t_{ij}$ fluctuate. 
Since the single-trace theory is a quadratic theory,
the integration over $\v{\phi}$ is not convergent
for general fluctuating source $t_{ij}$ in the absence of the quartic term.
This forces us to keep the quartic double-trace operator in the 
basis states as a regulator. 
However, the states labeled by $t_{ij}$ with a fixed $\lambda$ 
span the full symmetric Hilbert space.

For the generator of coarse graining transformation, 
we choose a Hamiltonian,
\bqa
\hat H = \sum_i \left[
 \frac{2}{m^2}  
  \v{\hat \pi}_i^\dagger \cdot  \v{\hat  \pi}_i 
+ i (  \v{\hat \phi}_i \cdot \v{\hat  \pi}_i + \v{ \hat \phi}^\dagger_i \cdot  \v{\hat \pi}^\dagger_i )
\right],
\eqa
where $\v{\hat \pi}_i, \v{\hat \pi}_i^\dagger$ are conjugate operators of $\v{\hat \phi}_i, \v{\hat \phi}_i^\dagger$
with the commutation relation
$
[ \hat \phi_{ia}, \hat \pi_{jb} ] = 
[ \hat \phi^\dagger_{ia}, \hat \pi^\dagger_{jb} ] = 
i \delta_{ij} \delta_{ab}$
with all other commutators being zero.
Here $a,b=1,2,..,N$ are flavor indices.
%%$
%%[ \hat \phi_{ia}, \hat \phi^\dagger_{jb} ] = 
%%[ \hat \pi_{ia}, \hat \pi^\dagger_{jb} ] = 
%%[ \hat \phi_{ia}, \hat \pi^\dagger_{jb} ] = 
%%[ \hat \phi^\dagger_{ia}, \hat \pi_{jb} ] = 0$.
%%
%%
$\cb S_0 \rb$ is the ground state of $\hat H^\dagger$ with energy $0$. 
This guarantees that the partition function is independent of $z$ in 
$Z = \lb S_0 \cb e^{-z \hat H} \cb t^{(0)} \rb$.
Multi-trace operators generated in 
each infinitesimal time evolution 
are removed such that 
\bqa
e^{-dz \hat H} \cb t^{(0)} \rb
= \int Dt^{(1)} ~~e^{ - N \sum_{ij} t^{*(1)}_{ij} ( t^{(1)}_{ij} - t^{(0)}_{ij} ) 
- dz N   {\cal H}[t^{*(1)}, t^{(0)}] }
\cb t^{(1)} \rb.
\eqa
Here ${\cal H}[t^*,t]$ is the coherent state representation of the quantum Hamiltonian,
\bqa
\hat {\cal H} & = &
\sum_i \left[ 
- \frac{2}{m^2} t_{ii}
+ ~ \frac{4\lambda\left(1+\frac{1}{N}\right)}{m^2} t^\dagger_{ii}
-4\lambda\left( t^\dagger_{ii} \right)^2
-\frac{8\lambda^2}{m^2}\left( t^\dagger_{ii} \right)^3  \right]  \nn
&& + \sum_{ij} \left[ 2    + \frac{4\lambda}{m^2} ( t^\dagger_{ii } +  t^\dagger_{jj} ) \right] t^\dagger_{ij}~ t_{ij} 
 - \frac{2}{m^2} \sum_{ijk} \left[ t^\dagger_{kj} t_{ki} t_{ij}  \right],
\label{eq:bulkH}
\eqa
where 
$t^\dagger_{ij}$ and $t_{ij}$ satisfy the commutation relation, 
$[t_{ij}, t^\dagger_{kl}] = \frac{1}{N} \delta_{ik} \delta_{jl}$.
$t_{ij}$ ($t_{ij}^\dagger$) is an operator 
that annihilates (creates) a connection between sites $i$ and $j$.
$t_{ij}$'s describe dynamical geometry of the $D$-dimensional space 
as the distance between two points is determined by the connectivity formed by the hopping fields.
This will become clear through an explicit derivation of metric from the hopping field in the following subsections.
For finite $N$, the strength of the connection is quantized.
Therefore, $\hat {\cal H}$ describes the evolution of the quantum geometry
under the RG flow.
The partition function is given by
\bqa
Z & =  & 
\lim_{z \rightarrow \infty}
\lb S_0 \cb e^{- {z}  \hat H } \cb t^{(0)} \rb 
=  
\left.
\int Dt(z)   ~~ \lb S_0 \cb t(\infty)  \rb  ~ 
e^{ -N  \int_0^{\infty}  dz \left(
 t^*_{ij} \partial_z t_{ij} 
+     {\cal H}[t^*,t] 
\right) }
\right|_{t_{ij}(0)=t^{(0)}_{ij} },
\label{eq:Zt}
\eqa
where $\int D t(z)$ sums over all RG paths
defined in the space of the single-trace operators.

As discussed below \eq{eq:C}, 
the bulk Hamiltonian is invariant under
a local transformation, $t_{ij} \rightarrow e^{i ( \varphi_i - \varphi_j)} t_{ij}$,
where $\varphi_i$ is site-dependent U(1) phase.
This symmetry originates from the U(1) gauge symmetry present in \eq{S1}
once the hopping parameters are promoted to dynamical fields.
The symmetry is broken only by the boundary condition in \eq{eq:Zt}
which fixes the dynamical hopping fields at $z=0$.
The symmetry forbids a bare kinetic term for the dynamical hopping fields
like $t_{ij}^\dagger t_{kl}$ in the Hamiltonian.
In other words, the bi-local object created by $t_{kl}^\dagger$
can not propagate to different links by itself. 
This does not mean that there is no propagating mode in the bulk.
Instead, this implies that there is no pre-determined background geometry
on which the bi-local objects propagate :
in order to specify kinetic term which involves gradient of fields, 
one needs to specify distances between points. 
In this case, the geometry is determined by the hopping fields themselves.
In order to find propagating modes in the bulk,
one first has to find the saddle-point configuration of the hopping fields.
Once the saddle point configuration is determined,
fluctuations of $t_{ij}$ around the saddle point can propagate 
on the geometry set by the saddle-point configuration.
This can be seen from the last term in \eq{eq:bulkH}
which gives rise to quadratic terms such as 
$  < t_{ij} > t^\dagger_{kj} t_{ki} $
once one of the field is replaced by the saddle point value.
Fluctuations of the dynamical sources propagate 
on the `shoulders' of their own condensates\cite{Lee2012}.
Equivalently, the fluctuations of the hopping field describes 
fluctuating background geometry.
This is similar to the situation in string theory, 
where perturbative strings propagate in the background
geometry set by condensate of strings.

In the large $N$ limit, one can use the saddle point approximation.
At the saddle point, $t^*_{ij}$ is not necessarily the complex conjugate of $t_{ij}$.
We denote the saddle point value of $t_{ij}$, $t^*_{ij}$ as 
$\bar t_{ij}$, $\bar p_{ij}$. 
The equations of motion for $ \bar t_{ij}$ and $\bar p_{ij}$ are
\bqa
\partial_z   \bar t_{ij}&=
-2 \Big\{
\frac{2\lambda\,\delta_{ij}}{m^2} 
- \delta_{ij} \left[4\lambda + \frac{12\lambda^2}{m^2}  \bar p_{ii} \right]  \bar p_{ii} 
+ \frac{2\lambda\, \delta_{ij}}{m^2} \sum_k \left( \bar t_{ik}  \bar p_{ik} +  \bar t_{ki}  \bar p_{ki} \right)  \nn
&\qquad \qquad \qquad \qquad \qquad + \left[
1+\frac{2\lambda}{m^2}\left(  \bar p_{ii}+ \bar p_{jj}\right)\right] \bar t_{ij}
-\frac{1}{m^2}\sum_{k}  \bar t_{ik}  \bar t_{kj} \Big\}, \nn
\partial_z  \bar p_{ij}&=2 \left\{
-\frac{\delta_{ij}}{m^2}
+\left[1+\frac{2\lambda}{m^2}\left(  \bar p_{ii}+  \bar p_{jj}\right)\right]  \bar p_{ij}
-\frac{1}{m^2}\sum_k\left(  \bar p_{ik}  \bar t_{jk}+ \bar t_{ki}  \bar p_{kj}\right)\right\}
\label{EOMs}
\eqa
with the two boundary conditions
\bqa
 \bar t_{ij}(0) & = & {t}^{(0)}_{ij},   \label{UV_BC}\\
 \bar p_{ij}(\infty) & = & \frac{1}{N} \left. \frac{\partial \ln \lb S_0 \cb   \bar t \rb }{\partial  \bar t_{ij}} \right|_{z=\infty}. \label{IR_BC}
\eqa

In \cite{Lunts2015}, the equations of motion were solved numerically for finite systems.
Although the numerical solution is reliable in the insulating states,
some of the results for gapless states are tainted with finite size effects. 
Here we solve the equations analytically in the thermodynamic limit.
We first note that $\v{\phi}$ at scale $z+dz$ is related to $\v{\phi}$ at scale $z$ through
the equation similar to \eq{OL},
$\v{\phi}_i(z+dz) = e^{- dz} \left( \v{\phi}_i(z) - \frac{\mu}{m} \sqrt{2 dz} \v{\Phi_i} \right)$,
where $\v{\Phi}_i$ is an auxiliary field with the action $\mu^2 | \v{\Phi}_i |^2$.
Because \eq{IR_BC} is satisfied at all $z$\cite{Lunts2015},
we have $ \bar p_{ij}(z+dz) = e^{-2dz} \left(  \bar p_{ij}(z) + \frac{2 dz}{m^2} \delta_{ij} \right)$.
This translates to the equation,
\bqa
\partial_z \bar p_{ij}(z) = -2 \bar p_{ij}(z) + \frac{2}{m^2} \delta_{ij}.
\label{pz}
\eqa
Now \eq{EOMs} and \eq{pz} are written in momentum space,
\bqa
\partial_z   \bar T_{q} &=&
-2 \Bigg\{
\frac{2\lambda}{m^2} 
-  \left[ 4\lambda + \frac{12\lambda^2}{m^2}  \bar p_{0} \right]  \bar p_{0} 
+ \frac{4\lambda}{V m^2} \sum_{q^{'}}  \bar T_{q^{'}}  \bar P_{q^{'}}  + 
\left[ 1+\frac{4\lambda}{m^2}   \bar p_{0} \right]  \bar T_{q}
-\frac{1}{m^2}   (\bar T_{q})^2 \Bigg\}, \label{EOMP1} \\
\partial_z  \bar P_{q} &=&
-\frac{2}{m^2}
+2 \left[1+\frac{4\lambda}{m^2}  \bar p_{0} \right]  \bar P_{q}
-\frac{4}{m^2}  \bar P_{q}   \bar T_{q},
\label{EOMP2} \\
\partial_z  \bar P_{q} &=&
-2 \bar P_q + \frac{2}{m^2}. \label{EOMP3}
\eqa
Here  $V$ is the volume of the $D$-dimensional space.
We assume the translational invariance,
and $ t_{ij}$ and $ p_{ij}$ depend only on $i-j$.
$ \bar p_0 \equiv  \bar p_{ii}$,
$ \bar T_q = \sum_r e^{i q r}  \bar t_{i+r i}$,
$ \bar P_q = \sum_r e^{-i q r}  \bar p_{i+r i}$.
Naively, it appears that  there are more equations than unknowns.
However, Eqs. (\ref{EOMP1})-(\ref{EOMP3}) are not over-determined
because only two of them are independent.
Combining \eq{EOMP2} and \eq{EOMP3}, one can solve $ \bar T_q(z)$, $\bar P_q(z)$
in terms of $\bar P_q(0)$,
\bqa
\bar T_q(z) &=&\frac{2\lambda}{m^2} + \frac{2 \lambda}{m^2} e^{-2z}( m^2  \bar p_0(0) - 1)
+ m^2 \frac{ m^2  \bar P_q(0) - 1 }{ m^2  \bar P_q(0) + e^{2z} - 1}, \label{eq:Tq} \\
\bar P_q(z) &=& e^{-2z} \bar P_q(0) + \frac{1}{m^2}( 1 - e^{-2z} ).
\label{eq:Pq}
\eqa
One can check that Eqs. (\ref{eq:Tq}) and (\ref{eq:Pq})
automatically satisfy \eq{EOMP1}.
By applying the UV boundary condition in \eq{UV_BC} to Eq. (\ref{eq:Tq}), 
$\bar  P_q(0)$ is determined to be
\bqa
 \bar P_q(0) & = & \frac{1}{ - \bar T_q(0) + m^2 + 2 \lambda  \bar p_0(0)   },
\label{Pq0}
\eqa
where $\bar P_q, \bar p_0$ satisfy the self-consistent equation
\bqa
 \bar p_0(0) & = & \frac{1}{V} \sum_q  \bar P_q(0).
\label{SCE}
\eqa
Eqs. (\ref{eq:Tq}) - (\ref{SCE}) completely determines $\bar T_q(z)$, $\bar P_q(z)$ from $\bar T_q(0)$.

\subsection{Zero density}
\label{zeromu}

First, we consider the case with zero chemical potential.
At the UV boundary, the form of $\bar T_q$ depends on
the choice of regularization (such as lattice type).
However, the locality and discrete lattice symmetries, if there is enough of them,
guarantee the universal form at low momentum,
$\bar  T_q(0) =  \bar T_0(0) - q^2 + O(q^4)$.
( For example, the cubic symmetry is enough to guarantee this form in three dimensions. )
Since we are mainly interested in the universal features that are independent of microscopic details,
we take $\bar  T_q(0) =  \bar T_0(0) - q^2$ to be valid at all momenta.
This amounts to using the continuum model with the two derivative kinetic term.
However, all the following discussions apply to theories with different regularizations at long distance limit
as far as the leading momentum dependence is $q^2$.
In the thermodynamic limit, 
\eq{SCE} can be converted to the momentum integral,
\bqa
 \bar p_0(0) & = & \frac{\bar P_0(0)}{V} + \int \frac{d^D q}{(2 \pi)^D}   \frac{1}{ q^2 + \delta^2  }.
\label{p00}
\eqa
Here $\delta^2 \equiv - \bar T_0(0) + m^2 + 2 \lambda  \bar p_0(0)$ is the gap.
$\frac{\bar P_0(0)}{V} $ is the condensate at zero momentum,
which needs to be singled out in the superfluid phase.
In the gapped phase, $\delta > 0$ and  $\frac{\bar P_0(0)}{V}= 0$.
The critical point is characterized by $\delta=0$ and $\frac{\bar P_0(0)}{V}= 0$.
In the superfluid phase, $\delta=0$ and $\frac{\bar P_0(0)}{V}> 0$.
The value of $\bar p_0(0)$ depends on the UV cut-off,
but its specific value is unimportant.
What matters is the physical mass gap, $\delta$.
We assume that $\bar T_0(0)$ is tuned so that $\delta$ is much smaller than the UV cut-off scale.
The dynamical hopping field 
and its conjugate field in the bulk become
\bqa
 \bar T_q(z) &=&
\frac{2\lambda}{m^2}  + m^2
+ \frac{2 \lambda}{m^2} e^{-2z}( m^2 \bar p_0(0) - 1)
- m^2 \frac{    \delta^2 +  q^2  }{  ( 1- e^{-2z} ) ( q^2 + \delta^2) + m^2 e^{-2z} }, \nn
\bar P_q(z) & = & \frac{e^{-2z}}{q^2 + \delta^2} + \frac{1-e^{-2z}}{m^2}.
\label{solution1}
\eqa
Although $\bar T_q(0)$ is quadratic in $q$,
$\bar T_q(z)$ with $z>0$ includes terms of  arbitrarily large powers of $q$.
In real space, this implies that the bi-local operators with arbitrarily large sizes
are generated along the RG flow as was indicated in Figs. \ref{fig:E1} and \ref{fig:E2}.
As will be shown in the next subsection, 
the hopping field has a direct relation with entanglement
contained in quantum state $e^{-z \hat H} \cb t^{(0)} \rb$.
The $z$ dependence of the long range hopping fields describe 
how entanglement spreads under the RG flow.  
In the presence of the bi-local fields with all sizes,
the tensor network in the bulk may or may not be local
depending on $\delta$ and $z$.

From the saddle point solution in \eq{solution1}, 
one can compute the metric in the bulk.
In order to extract the metric,
it is useful to consider fluctuations of the hopping fields around the saddle point,
\bqa
\td t_{ij} = t_{ij} - \bar t_{ij}, ~~~
\td p_{ij} = t_{ij}^* - \bar p_{ij}.
\eqa
The quadratic action for the fluctuations is given by
\bqa
S_2 & = & \int dz \Bigg\{
-4 \lambda \sum_i 
\left( 1+ \frac{6 \lambda}{m^2} \bar p_0 \right) \td p_{ii}^2 \nn
&& +  \sum_{ij} \left[ 
\td p_{ij} \left( \partial_z  + 2 + \frac{8 \lambda}{m^2} \bar p_0 \right) \td t_{ij} 
 + \frac{4 \lambda}{m^2}
(\td p_{ii} + \td p_{jj} ) 
\left( 
\bar p_{ij} \td t_{ij}
+ \bar t_{ij} \td p_{ij}
\right) \right] \nn
&& - \frac{2}{m^2} \sum_{ijk} 
\left(
\bar p_{ij} \td t_{ik} \td t_{kj}
+  \bar t_{ik}  \td p_{ij} \td t_{kj}
+  \bar t_{kj}  \td p_{ij} \td t_{ik}
\right)
\Bigg\}.
\label{eq:bulkaction}
\eqa
Here $\td p_{ij}$ is the conjugate momentum of $\td t_{ij}$.
The action includes terms that are quadratic in the momentum.
The last two terms in \eq{eq:bulkaction} describes the kinetic term
for the bi-local field that is generated from the cubic interaction in \eq{eq:bulkH}.
For example, $  \bar t_{ik}  \td p_{ij} \td t_{kj} $ describes the process
where one end of the bi-local field moves from site $k$ to $i$
as if a man takes a step with one foot while the other foot pivoted on the ground.
The range of step is not pre-fixed, 
but is dynamically determined by how fast the saddle point configuration $\bar t_{ik}$ decays in $|i-k|$.
Therefore, the locality in the bulk is a dynamical feature
which is determined by length scale in $\bar t_{ik}$.
In the partition function, 
the paths of $\td t_{ij}$ 
and $\td p_{ij}$ in the complex plane 
need to be chosen 
along the direction of the steepest descent.

The equations of motion for $\td t_{ij}, \td p_{ij}$ are
\bqa
&& \left( \partial_z + 2 + \frac{ 8 \lambda}{m^2} \bar p_0(z) \right) \td t_{ij}
 - \frac{2}{m^2} \left( 
\bar t_{ik} \td t_{kj}  + \bar t_{kj} \td t_{ik} \right) 
+ \frac{4 \lambda}{m^2} \bar t_{ij} ( \td p_{ii} + \td p_{jj} )
\nn
&& + \delta_{ij} \left[
-8 \lambda \left( 1+ \frac{6 \lambda}{m^2} \bar p_0(z) \right) \td p_{ii}
+ \frac{4 \lambda}{m^2} \left(
\bar p_{ik}( \td t_{ik} + \td t_{ki} )
+ \bar t_{ik} ( \td p_{ik} + \td p_{ki} ) 
\right) 
\right] 
 = 0, 
\label{QEOM1} \\
 && \left( - \partial_z + 2 + \frac{ 8 \lambda}{m^2} \bar p_0(z) \right) \td p_{ij}
+ \frac{4 \lambda}{m^2} \bar p_{ij} ( \td p_{ii} + \td p_{jj} ) 
 - \frac{2}{m^2} \left( 
\bar p_{ik} \td t_{jk} + \bar p_{kj} \td t_{ki} 
+ \bar t_{ki} \td p_{kj}  + \bar t_{jk} \td p_{ik} \right) =0. \label{QEOM2} \nn
\eqa
Once the equation of motion for $\td p_{ij}$ is solved,
the equations for $\td t_{ij}$ in general include second order
derivatives in $z$ due to the quadratic kinetic term in \eq{eq:bulkaction}.
However, we don't have to consider the full equations of motion to extract the geometry.
This is because $\td t_{ij}$ is a probe that propagates on the geometry set by $\bar t_{ij}$, 
and the background geometry is independent of dynamics of the specific mode.
The equation for the anti-symmetric part of the hopping field 
defined by $\td t^A_{ij} = \td t_{ij} - \td t_{ji}$ satisfies a simpler equation of motion,
\bqa
\left( \partial_z + 2 + \frac{ 8 \lambda}{m^2} \bar p_0(z) \right) \td t^A_{ij}
 - \frac{2}{m^2} \left( 
\bar t_{ik} \td t^A_{kj}  + \bar t_{kj} \td t^A_{ik} \right)  = 0.
\label{QEOMA} 
\eqa
The  conjugate momentum decouples from the equation of $\td t^A_{ij}$
because the saddle point configuration is symmetric.
As a result, the anti-symmetric mode propagates  diffusively in the bulk rather than ballistically.
Although $\bar t_{ij} = \bar t_{ji}$ at the saddle point without background gauge field,
$t_{ij}$ is independent of $t_{ji}$ as a dynamical field.
This is a difference of the $U(N)$ model from the $O(N)$ model. 
Therefore, $t_{ij}$ includes both even and odd spin fields in the expansion,
\bqa
t_{ij} = \sum_{n=0}^\infty
\frac{1}{n!}
 \sum_{\mu_1,\mu_2,..,\mu_n=0}^{D-1} 
(x_i-x_j)^{\mu_1}
(x_i-x_j)^{\mu_2}
..
(x_i-x_j)^{\mu_n}
~
t_{\mu_1,\mu_2,..,\mu_n}\left( \frac{x_i + x_j}{2} \right),
\eqa 
where $x_i$ is $D$-dimensional coordinate associated with site $i$.
$\td t_{ij}^A$ describes the fluctuations in the odd spin sector.
In the following, we compute the bulk metric 
in the gapped and the gapless states respectively
using \eq{solution1} and \eq{QEOMA}.

\subsubsection{Gapped state}

\begin{figure}[h]
\centering
\includegraphics[width = 0.65\columnwidth]{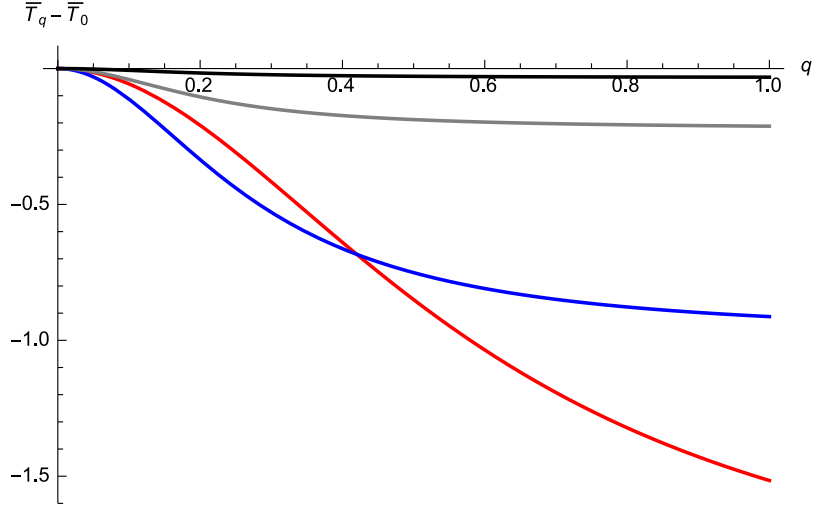}
\caption{
Plot of $\bar T_q(z)-\bar T_0(z)$ 
as a function of $q$ for $z=1,2,3,4$ in the order of increasing flatness
in the gapped phase with $\delta=0.2, m^2=2, \lambda=1$.
}
\label{fig:Tp_gap}
\end{figure}

In \fig{fig:Tp_gap}, $\bar T_q(z)$ in \eq{solution1} is plotted as a function of momentum at different radial slices in the gapped phase.
$\bar T_q(z)$ is peaked around $q=0$, but 
the dispersion goes to zero exponentially in $z$,
\bqa
\bar T_q(z) = \bar T_0(z) - \frac{ m^4}{\delta^4} e^{-2z} q^2 + e^{-2z} O(q^4).
\eqa
$\bar T_q(z)$ stays analytic in $q$ at all $z$.
In the large $z$ limit, $\bar T_q(z)$ becomes completely flat.
In the gapped phase, $e^{-z \hat H} \cb  t^{(0)} \rb$ is smoothly projected to the direct product state under RG.
In real space, the non on-site hopping field becomes
\bqa
%\lim_{z \rightarrow \infty}  t_{ij}(z) = \frac{2\lambda}{m^2}  \delta_{ij} + O(e^{-2z}) .
\bar t_{ij}(z)  & \sim &  e^{-2z} \frac{1}{ |i-j|^{D-2} } ~~~\mbox{ for $|i-j| < \xi(z)$}, \nn
& \sim &  e^{-2z} e^{- |i-j| / \xi(z)} ~~~\mbox{ for $|i-j| > \xi(z)$},
\label{t_decay}
\eqa 
where the range of hopping is given by 
\bqa
\xi(z) = ( \delta^2 + m^2 e^{-2z} )^{-1/2}.
\label{eq:xi}
\eqa 
As we will see in the following, 
\eq{eq:xi} is the length scale 
that controls the range of entanglement 
and the geometry in the bulk.

In order  to read the geometry, 
we first rewrite \eq{QEOMA} in the Fourier space,
\bqa
\left( \partial_z + 2 + \frac{ 8 \lambda}{m^2} \bar p_0(z) 
 - \frac{2}{m^2} \left[ \bar T_{p_1} + \bar T_{p_2}  \right]  
\right) \td T^A_{p_1, p_2}
= 0,
\label{EOMAF}
\eqa
where $\td T^A_{p_1, p_2} = \sum_{ij} e^{i ( p_1 x_i - p_2 x_j )} \td t^A_{ij}$,
and $\bar T_q$ in \eq{solution1} is written as
\bqa
\bar T_q(z) = 2 \lambda \bar p_0(z) - m^2 e^{-2z} + m^2 
A(z)
 \frac{ 1  }{ 1 +  B(z)     \frac{ q^2  }{m^2} }
\eqa
to the leading order in $e^{-z}$, where 
$A(z) = ( e^{-z} m  \xi(z) )^2$,
$B(z) = ( m \xi(z) )^2$.
%%$A(z) = \frac{1}{1 +   \left( \frac{ \delta}{m}  e^{z} \right)^2 }$,
%%$B(z) = \frac{1}{ \left( \frac{ \delta}{m} \right)^2 + e^{-2z} } $.
Here $B(z)$ sets the scale factor for the derivatives in the $D$-dimensional space to all orders,
and it becomes the metric component $g^{00}=g^{11}=..=g^{D-1,D-1}$.
It is noted that the range of hopping in \eq{eq:xi} determines the unit proper distance 
: two points separated by $\xi(z)$ sites are considered to be within a unit proper distance (in a fixed scale $1/m$) 
because you can jump by $\xi(z)$ sites in one hopping.
On the other hand, $A(z)$ sets the relative scale factor between
the $z$-derivative and the derivatives in the $D$-dimensional space.  
As a result, $A(z)^{-1}$ should be identified as the metric component in the $z$ direction.
By multiplying $m^2 A(z)^{-1}$ to \eq{EOMAF},
the equation for the continuum field 
defined by $\td t^A(x,x^{'},z)= e^{2z} ~\int dp_1 dp_2 ~ e^{- i (p_1 x - p_2 x^{'} )} \td T^A_{p_1,p_2}(z)$
can be expressed in the covariant form,
\bqa
 \left(  
m  \sqrt{g^{zz}}  \partial_z
 + 4 \delta^2  
- \frac{ 2 m^2  }{ 1 - \frac{ g^{\mu \nu} \partial_\mu \partial_\nu  }{m^2} }
- \frac{ 2 m^2 }{ 1 - \frac{ g^{\mu \nu} \partial_\mu^{'} \partial_\nu^{'}  }{m^2} }
\right) \td t^A(x,x^{'}, z) = 0,
\label{ce0}
\eqa
where
$g^{zz}  = m^2 A(z)^{-2}$,
$g^{00}=g^{11}=..=g^{D-1,D-1}  =   B(z)$.
\eq{ce0} describes the bi-local field that propagates diffusively 
with the diffusion coefficient $D_0 \sim 1/m$
in a curved space\cite{2012NJPh...14b3019S} with the metric,
\bqa
ds^2 = 
 \left(\frac{1}{ 1 +   \left( \frac{ \delta}{m}  e^{z} \right)^2 } \right)^2 
\frac{dz^2}{m^2} +
 \left( \left( \frac{ \delta}{m} \right)^2 + e^{-2z} \right) 
  \sum_{\mu=0}^{D-1} dx^\mu dx^\mu.
  \label{eq:metric}
\eqa

For $\delta \neq 0$, $g_{zz}$ decays exponentially in the large $z$ limit.
Since $e^{-z \hat H} \cb t^{(0)} \rb$ is already close to the direct product state at large $z$,
an additional application of the projection operator $e^{-dz \hat H}$ 
adds only an exponentially small proper distance in the radial direction.
This can be understood more precisely
in terms of the entanglement entropy of  $e^{-z \hat H} \cb t^{(0)} \rb$.
In the large $z$ limit, 
the von Neumann entanglement entropy between
region $A$ and its complement is given by
\bqa
S_E = N \sum_{i,j}^{'} | \bar t_{ij}(z) |^2 G_0^2 \left[  1 - \ln ( |\bar t_{ij}(z)|^2 G_0^2 )  \right] + O( \bar t^3 )
\label{eq:EE}
\eqa
in the large $N$ limit 
(See Appendix A for the derivation).
Here $\sum_{i,j}^{'}$ sums over ordered pairs of sites
which are spanned across $A$ and $\bar A$.
$G_0$ is the on-site propagator defined in \eq{eq:G0}.
Because the entanglement is directly related to the hopping fields,
the spread of the entanglement is controlled by the same length scale in \eq{eq:xi}
which determines the metric.
Equivalently, $\xi(z)$ controls the range within (outside of) which two sites have algebraically (exponentially) decaying mutual information. 
Therefore, entanglement entropy of a subsystem with linear size smaller (greater) than $\xi(z)$
exhibits an super-area (area) behavior.

The entanglement entropy of a single lattice site with the rest of the system is given by
$s_E   \sim N e^{-4z} \xi(z)^{4-D}$ upto a logarithmic correction for $D<4$. 
Here \eq{t_decay} and \eq{eq:xi} are used.
However, $s_E$ is not a good measure of the overall entanglement because 
a single site does not corresponds to the unit physical volume 
measured in the metric in \eq{eq:metric}.
Therefore, we consider the entanglement entropy of a region 
centered at the origin with unit proper volume in $D$ dimensions,
\bqa
s_E^{'} =   N \sum_{
\{ |i| < \xi(z) <  |j| \}
\cup
\{ |i| > \xi(z) >  |j| \}
} | \bar t_{ij}(z) |^2 G_0^2 \left[  1 - \ln ( |\bar t_{ij}(z)|^2 G_0^2 )  \right]
\sim N e^{-4z} \xi(z)^4,
\label{sE2}
\eqa
where $i$, $j$ are divided by $\xi(z)$ because
coordinate distance $\xi(z)$ corresponds to the fixed proper distance $1/m$.
$s_E^{'}$ measures the entanglement of the largest possible region
within which the entanglement entropy exhibits a super-area behavior.   
In the gapped phase, $\xi(z)$ approaches $1/\delta$ in the large $z$ limit.
As a result, $s_E^{'}$ decays as $e^{-4z}$ in $z$.
This explains why the radial metric decays exponentially in $z$.
In contrary, $s_E^{'}$ does not go to zero in gapless states 
as will be discussed in the next section.

The proper distance from the UV boundary ($z=0$) 
to the IR limit ($z=\infty$) in the unit of $m$ is given by
\bqa
\tau =   m \int_0^\infty dz \sqrt{g_{zz}} = \frac{1}{2} \ln \left(  1+ \left( \frac{m}{\delta} \right)^2 \right).
\label{eq:R}
\eqa
In the gapped phase, $\tau$ is finite. 
The bulk geometry smoothly ends
at a finite depth in the radial direction.
From field theory point of view, $\tau$ measures the proper RG time
that is needed for a theory to flow to the IR fixed point.
In this sense, the geometric distance $\tau$ in the bulk 
provides a notion of distance in the space of field theories.
Alternatively, $\tau$  measures the complexity of the UV state : 
the depth of the RG transformation that is needed to remove all entanglement 
present in the UV state\cite{2014arXiv1402.5674S}.
This demonstrates the connection between 
entanglement and geometry\cite{
PhysRevLett.96.181602,
1126-6708-2007-07-062,
VanRaamsdonk:2010pw,
Casini2011,
Lewkowycz2013}
from the first principle construction.

As the system is tuned toward the critical point, $\delta$ become smaller.
In the small $\delta$ limit, $\tau$ diverges logarithmically.
This implies that the critical point can not be projected to the insulating fixed point
by RG transformations with a finite depth.

%%%%%%%%%%
%%%%%%%%%%

\subsubsection{Gapless states}

\begin{figure}[h]
\centering
\includegraphics[width = 0.65\columnwidth]{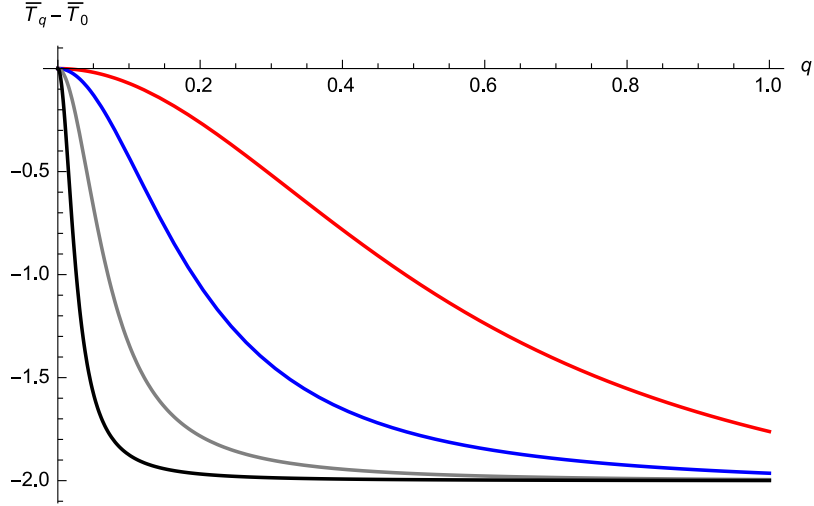}
\caption{
Plot of $\bar T_q(z)-\bar T_0(z)$ at the critical point 
as a function of $q$ for $z=1,2,3,4$ from top to bottom
with $\delta=0, m^2=2, \lambda=1$. 
With increasing $z$, the peak at $q=0$ gets sharper.
}
\label{fig:Tp}
\end{figure}

Now we turn to the critical point with $\delta = 0$.
$\bar T_q(z)$ in \eq{solution1} is shown at the critical point in \fig{fig:Tp}.
As $z$ increases, $\bar T_q(z)$ becomes flatter at nonzero $q$ 
as the short distance entanglement is removed by the projection operator.
However,  the peak at $q=0$ never disappears at any $z$ unlike in the gapped phase.
In the large $z$ limit, the peak becomes infinitely sharp,
and $\bar T_q(z)$ becomes non-analytic at $q=0$.
In other words, $z \rightarrow \infty$ and $q \rightarrow 0$ limits do not commute.
If one takes the small $q$ limit first at a large but finite $z$, one obtains
\bqa
 \bar T_q(z) = \bar T_0(z) 
 %%\left(  m^2 + \frac{ 2 \lambda}{m^2} ( 1 - e^{-2z} ) + 2 \lambda  \bar p_0(0) e^{-2z} \right)
-  e^{2z} q^2 + \frac{(  e^{2z} q^2)^2}{m^2} + ...
\eqa  
The coefficients of the $q$ dependent terms grow
indefinitely with increasing $z$.  
This implies that $\bar t_{ij}$ becomes non-local in the large $z$  limit
as can be seen by setting $\delta=0$ in \eq{eq:xi}.
The overall amplitude of the hopping field
is exponentially small in the large $z$ limit.
However, the height of the peak at $q=0$ remains order of $1$ 
because the range of hopping also diverges exponentially in $z$.

In the gapless phase, $\xi(z)$ diverges as $e^{z}/m$ in the large $z$ limit.
As a result, the entanglement within a unit proper volume in \eq{sE2} does not vanish in the large $z$ limit.
Although the entanglement of individual site decreases exponentially with increasing $z$,
the entanglement spreads out to larger region that
the entanglement within unit proper volume approaches a non-zero constant in the large $z$ limit.
The persistent entanglement is responsible for the divergent proper distance in the radial direction.
The fluctuations of the hopping field satisfy the same equation in \eq{ce0}.  
With $\delta=0$, \eq{eq:metric} is reduced to the $AdS_{D+1}$ metric,
\bqa
ds^2 = \frac{dz^2}{m^2} + e^{-2z} \sum_{\mu=0}^{D-1} dx^\mu dx^\mu.
\label{ADSmetric}
\eqa
The curvature length scale of the bulk space is $1/m$.
Higher derivative terms in \eq{ce0} are suppressed in the same scale.
The extended IR geometry of the anti-de Sitter space 
is due to the inability of removing all 
entanglement in the UV state through a 
projection operator with a finite depth.
In this sense, the gapless state is infinitely far away from the insulating fixed point.

In the superfluid phase, 
the condensate  in \eq{p00} modifies the value of $ p_0(0)$.
However, $\bar T_q(z)$ is still given by the same expression in \eq{solution1} with $\delta=0$. 
In \cite{Lunts2015}, based on numerical solutions in finite system sizes, 
it was argued that the hopping field exhibits 
a divergent length scale at finite $z$ in the superfluid phase.
However, this claim is incorrect : 
the apparent divergence `observed' in the numerical solution is a finite size effect.
The exact expression in \eq{solution1} shows that
the range of hopping diverges only in the large $z$ limit
in the thermodynamic limit.
Therefore the bulk geometry probed by the fluctuation field
is given by the same $AdS_{D+1}$ space.

When the system is in the `wrong' phases (critical point or superfluid) 
compared to the guessed IR fixed point (insulating state),
the UV state can not be smoothly projected to the IR state.
This obstruction culminates in the non-analyticity of $\bar T_q(z)$  in the large $z$ limit.
The non-analytic behavior of the hopping field in the large $z$ limit
can be viewed as a critical behavior associated with a dynamical phase transition\cite{PhysRevLett.110.135704,PhysRevLett.115.140602}.
The divergent length scale in the spread of entanglement gives rise to the Poincare horizon at $z=\infty$.
We use the term `phase transition' in a more general context 
despite the fact that we can not reach the other side of the critical point in this particular case.
In the bosonic vector model, the critical point lies at $z=\infty$
because the gapless states have no scale.
Even in the superfluid phase, order of $N$ Goldstone modes 
are described by the scale invariant theory.  
However, a horizon can arise at finite $z$ in gapless states which possess intrinsic scales. 
For example, a dynamical phase transition can occur at finite $z$ in metallic states in fermionic models\cite{Hu}.

The geometric obstruction to smooth projection between two phases exists only in the thermodynamic limit.
For finite systems, there is always a nonzero gap due to the finite size effect, and $\bar T_q(z)$ is analytic in $q$ at all $z$.
In this case, any state can be smoothly projected to the direct product state. 
This is in line with the fact that distinct phases are sharply defined
only in the thermodynamic limit.

\subsection{Finite density}
\label{nonzeromu}

Now we consider the case with a non-zero chemical potential.
In the $D$-dimensional Euclidean lattice,
we pick one of the direction to be an imaginary time.
A chemical potential enters in the hopping field
 as an imaginary gauge potential along the Euclidean time direction.
With a chemical potential $\mu$, 
the hopping field is modified as
\bqa
\bar T_q(0) = \bar T_0(0) - (q_0 - i \mu)^2 - |\vec q|^2
\eqa
at the UV boundary, 
where $\vec q = (q_1, q_2, .., q_{D-1})$ represents 
the $(D-1)$-dimensional momentum.
The hopping field in the bulk becomes
\bqa
 \bar T_q(z) &=&
\frac{2\lambda}{m^2}  + m^2
+ \frac{2 \lambda}{m^2} e^{-2z}( m^2  p_0(0) - 1)
- m^2 \frac{    \delta^2 +  (q_0 - i \mu)^2 + |\vec q|^2 }{  ( 1- e^{-2z} ) \left( (q_0 - i \mu)^2 + |\vec q|^2 + \delta^2 \right) + m^2 e^{-2z} }. \nn
\label{solution2}
\eqa

If $\delta >  \mu$, the gap remains non-zero,
and the system is in the insulating state. 
The behaviour of $\bar t_{ij}$ in the bulk is essentially the same
as the one in the insulating state at zero chemical potential.
The hopping field remains analytic in momentum at all $z$.
In the large $z$ limit, $\bar t_{ij}$ becomes on-site term in real space, 
and the bulk space smoothly ends at a finite depth in the radial direction.

At the critical point and in the superfluid phase, 
we have $\delta = \mu$.
%%The difference of the superfluid phase from the critical point is 
%%that there is a condensate at zero momentum,
%%which modifies the equation for $\bar p_0$ as
%%
%%$\bar p_0(0)  =  \frac{1}{V}  \bar P_0(0)
%%+ \int \frac{d^D q}{(2 \pi)^D}   \frac{1}{ -2 i \mu q_0 + q_0^2 + |\vec  q|^2 }$.
%%
The hopping field $\bar T_q(z)$ is given by
\bqa
 \bar T_q(z) &=&
\frac{2\lambda}{m^2}  + m^2
-  m^2 \frac{ \frac{e^{2z}}{m^2} ( -2i \mu q_0 + q_0^2 + |\vec q|^2 )   }
{ 1 + \frac{e^{2z}}{m^2} ( -2i \mu q_0 + q_0^2 + |\vec q|^2 )    }
\label{solution4}
\eqa
in the large $z$ limit.
Similar to the charge neutral case, 
$\bar T_q(z)$ becomes singular at $q=0$ 
in the large $z$ limit.
The difference is that  $q_0$-linear term is dominant at low frequencies.

To extract the metric in the bulk, 
we again consider fluctuations around the saddle point, $\td t_{ij} =  t_{ij} - \bar t_{ij}$.
With a non-zero chemical potential, $\bar t_{ij}$ is no longer identical to $\bar t_{ji}$.
Since the anti-symmetric component is not decoupled from the symmetric component,
\eq{QEOMA}  does not hold.
However, \eq{QEOM1} is still simplified in the large $|i-j|$ limit,
where only the first two terms are important.
The continuum field  defined by 
$\td t(x, x^{'}, z) = e^{2z} \td t_{ij}(z)  $
satisfies 
\bqa
\left(  
m^2  \partial_z 
- \frac{ 2m^2 }{ 1 -  \frac{1}{m^2} \left( 2 \mu \sqrt{g^{00}}  \partial_0 + g^{\alpha \beta} \partial_\alpha \partial_\beta \right) }
- \frac{ 2m^2 }{ 1 - 
\frac{1}{m^2} \left( 2 \mu \sqrt{g^{00}}  \partial_0^{'} + g^{\alpha \beta} \partial_\alpha^{'} \partial_\beta^{'} \right) }
\right) \td t(x, x^{'}, z) = 0 \nn
\label{ce2}
\eqa
in the large $|x-x^{'}|$ limit,
where
$g^{00} = e^{4z}$,
$g^{11}=g^{22}=..=g^{D-1,D-1} = e^{2z}$,
and  $\alpha,\beta$ runs from $1$ to $D-1$.
Therefore the fluctuation fields propagate in the Lifshitz background\cite{PhysRevD.78.106005,Balasubramanian2010,2010JHEP...12..002D} 
whose metric is given by
\bqa
ds^2 =  \frac{dz^2}{m^2} + e^{-4z}  (dx^0)^2  + e^{-2z}   d \vec x^2.
\eqa
The geometry is invariant under the anisotropic scaling with 
the dynamical critical exponent  $2$,
\bqa
z \rightarrow z+ s, 
~~~ x^0 \rightarrow  s^{2} x^0, 
~~~ \vec x \rightarrow s \vec x.
\eqa
The non-locality of the hopping amplitude in the large $z$ limit
gives rise to the horizon at $z=\infty$.

\section{ Horizon from dynamical phase transition  }
\label{DPT}

The renormalization group flow in Euclidean field theory
is a gradual collapse of a quantum state 
associated with an action $S_1$,
\bqa
e^{-z \hat H} \cb S_1 \rb,
\label{EVOL}
\eqa
where $ S_1$ is the deformation turned on at an IR fixed point $S_0$,
and the Hamiltonian $\hat H$ is the generator of coarse graining
whose ground state is given by the IR fixed point.
If $z$ is viewed as an imaginary time (which is different from $x_0$ along the field theory direction),
\eq{EVOL} describes a time-dependent quantum state.
Under the evolution, 
the state is gradually projected  
to the ground state of $\hat H$.
For systems with a finite number of degrees of freedom,
any initial state is projected to the ground state
smoothly.
In the thermodynamic limit, however, 
a smooth projection is not always possible.
Whether $\cb S_1 \rb$ can be smoothly projected to the ground state of $\hat H$
depends on whether the system described by the full action $S_0 + S_1$
is in the same phase described by the assumed IR fixed point.

In the examples considered in the previous section,
$\cb S_0 \rb$ is chosen to be a direct product state,
which is the fixed point theory for the insulator.
The Hamiltonian gradually removes short-range entanglement 
to project the UV state to the direct product state.
At the same time, short-range entanglement is transferred to larger scales
as long-range tensors are generated in the tensor product representation of \eq{EVOL}. 
When small deformations (such as hopping or chemical potential) are turned on,
the system is still in the insulating phase because
small deformations are irrelevant at the insulating fixed point. 
In this case, the spread of entanglement is confined within a finite length scale, 
which is controlled by the gap of the system.
The UV state smoothly evolves to the direct product state,
and the geometry in the bulk ends at a finite proper distance in the radial direction.
The proper distance from the UV boundary to the IR limit gives
the depth of RG transformation that is required to remove all entanglement in the UV state.
The first principle derivation of the metric in the bulk provides a physical meaning for the geometry in the bulk. 
The metric in the field theory directions measures the range over which entanglement spreads in \eq{EVOL},
while the metric in the radial direction is proportional to the amount of entanglement present 
in the region within which the entanglement entropy exhibits a super-area behavior.

As the deformations get larger, 
it takes longer RG time for the state to be projected to the direct product state.
If the deformations are stronger than certain critical strength, 
the system is no longer in the insulating phase.
In this case, entanglement spreads to arbitrarily large length scale under the RG flow.
At the same time, the entanglement contained within a proper volume does not decay to zero.  
The persistent entanglement gives rise to an extended geometry in the IR : 
the anti de-Sitter space for the charge neutral case
and the Lifshitz geometry for the charged case.   
In the large $z$ limit, \eq{EVOL} develops a singularity
due to the divergent length scale associated with the spread of entanglement. 
The Poincare horizon that emerges in the large $z$ limit
corresponds to the critical point associated with a global spread of entanglement.
The on-set of non-locality shows up as horizon in the bulk geometry.

One possible use of viewing horizon as critical point is to explain its universality.
As critical points are characterized by a small number of universal exponents
which are independent of microscopic details,
different types of horizons are characterized by macroscopic parameters.
Under the present picture, these two are the same thing.
The extreme red shift present near horizon can be viewed as the critical slow down near critical point.

In order to obtain a bulk with the Lorentzian signature,
a natural way would be to generate collapse of wave function 
not from $e^{-z \hat H}$ but from $e^{-i z \hat H}$.
Although the latter does not change the norm of states,
it creates a precession that effectively
projects out fast modes to observers with finite resolution.
Depending on the sign of the kinetic term in the bulk Hamiltonian, which originates from the beta functions of multi-trace operators,
the radial direction can be either space-like or time-like\cite{Lee:2013dln}.
It will be of interest to see how the Lorentzian anti-de Sitter space
or the de-Sitter space can be obtained from concrete field theories\cite{2001JHEP...10..034S, 2003AnPhy.303...59B, 2011arXiv1108.5735A}.

 \section{Summary }
\label{summary}

In summary, we showed that 
RG flow in Euclidean field theories can be
understood as a gradual wave function collapse.
Once the final state is chosen to be a direct product state, 
the RG transformation is generated by a quantum Hamiltonian
that gradually removes short-range entanglement in the UV state 
which originates from the action of field theory.
When the system is in the gapped phase described by the final state,
the UV state is smoothly projected to the direct product state.
In this case, the bulk geometry smoothly ends at a finite proper distance in the radial direction.
The geometric distance in the radial direction measures a complexity of the UV state :
the depth of RG transformation needed to remove all entanglement.
On the other hand, the proper distance in the field theory direction is
determined by the spread of entanglement.
Unlike gapped states, gapless states can not be smoothly projected to the direct product state
due to the spread of entanglement to arbitrarily long distance scales under RG flow.
The impossibility of projecting the UV state to the IR state with the projector of finite depth
gives rise to the extended geometry in the bulk.
In the long RG time limit, the scale dependent state develops a singularity, exhibiting a critical behavior.
The critical point gives rise to the horizon in the bulk.

\section{Acknowledgments}
The author thanks the participants of the 
 Simons symposiums on quantum entanglement
and the Aspen workshop on emergent spacetime in string theory
for inspiring discussions.
The research was supported in part by 
the Natural Sciences and Engineering Research Council of 
Canada, the Early Research Award from the Ontario Ministry of 
Research and Innovation, and
the Templeton Foundation.
Research at the Perimeter Institute is supported 
in part by the Government of Canada 
through Industry Canada, 
and by the Province of Ontario through the
Ministry of Research and Information.

\bibliography{references}

\begin{thebibliography}{59}
\expandafter\ifx\csname natexlab\endcsname\relax\def\natexlab#1{#1}\fi
\expandafter\ifx\csname bibnamefont\endcsname\relax
  \def\bibnamefont#1{#1}\fi
\expandafter\ifx\csname bibfnamefont\endcsname\relax
  \def\bibfnamefont#1{#1}\fi
\expandafter\ifx\csname citenamefont\endcsname\relax
  \def\citenamefont#1{#1}\fi
\expandafter\ifx\csname url\endcsname\relax
  \def\url#1{\texttt{#1}}\fi
\expandafter\ifx\csname urlprefix\endcsname\relax\def\urlprefix{URL }\fi
\providecommand{\bibinfo}[2]{#2}
\providecommand{\eprint}[2][]{\url{#2}}

\bibitem[{\citenamefont{Maldacena}(1999)}]{Maldacena:1997re}
\bibinfo{author}{\bibfnamefont{J.~M.} \bibnamefont{Maldacena}},
  \bibinfo{journal}{Int.J.Theor.Phys.} \textbf{\bibinfo{volume}{38}},
  \bibinfo{pages}{1113} (\bibinfo{year}{1999}), \eprint{hep-th/9711200}.

\bibitem[{\citenamefont{Witten}(1998)}]{Witten:1998qj}
\bibinfo{author}{\bibfnamefont{E.}~\bibnamefont{Witten}},
  \bibinfo{journal}{Adv.Theor.Math.Phys.} \textbf{\bibinfo{volume}{2}},
  \bibinfo{pages}{253} (\bibinfo{year}{1998}), \eprint{hep-th/9802150}.

\bibitem[{\citenamefont{Gubser et~al.}(1998)\citenamefont{Gubser, Klebanov, and
  Polyakov}}]{Gubser:1998bc}
\bibinfo{author}{\bibfnamefont{S.}~\bibnamefont{Gubser}},
  \bibinfo{author}{\bibfnamefont{I.~R.} \bibnamefont{Klebanov}},
  \bibnamefont{and} \bibinfo{author}{\bibfnamefont{A.~M.}
  \bibnamefont{Polyakov}}, \bibinfo{journal}{Phys.Lett.}
  \textbf{\bibinfo{volume}{B428}}, \bibinfo{pages}{105} (\bibinfo{year}{1998}),
  \eprint{hep-th/9802109}.

\bibitem[{\citenamefont{{Akhmedov}}(1998)}]{1998PhLB..442..152A}
\bibinfo{author}{\bibfnamefont{E.~T.} \bibnamefont{{Akhmedov}}},
  \bibinfo{journal}{Physics Letters B} \textbf{\bibinfo{volume}{442}},
  \bibinfo{pages}{152} (\bibinfo{year}{1998}), \eprint{hep-th/9806217}.

\bibitem[{\citenamefont{de~Boer et~al.}(2000)\citenamefont{de~Boer, Verlinde,
  and Verlinde}}]{deBoer:1999xf}
\bibinfo{author}{\bibfnamefont{J.}~\bibnamefont{de~Boer}},
  \bibinfo{author}{\bibfnamefont{E.~P.} \bibnamefont{Verlinde}},
  \bibnamefont{and} \bibinfo{author}{\bibfnamefont{H.~L.}
  \bibnamefont{Verlinde}}, \bibinfo{journal}{JHEP}
  \textbf{\bibinfo{volume}{0008}}, \bibinfo{pages}{003} (\bibinfo{year}{2000}),
  \eprint{hep-th/9912012}.

\bibitem[{\citenamefont{Skenderis}(2002)}]{Skenderis:2002wp}
\bibinfo{author}{\bibfnamefont{K.}~\bibnamefont{Skenderis}},
  \bibinfo{journal}{Class.Quant.Grav.} \textbf{\bibinfo{volume}{19}},
  \bibinfo{pages}{5849} (\bibinfo{year}{2002}), \eprint{hep-th/0209067}.

\bibitem[{\citenamefont{Heemskerk and Polchinski}(2011)}]{Heemskerk:2010hk}
\bibinfo{author}{\bibfnamefont{I.}~\bibnamefont{Heemskerk}} \bibnamefont{and}
  \bibinfo{author}{\bibfnamefont{J.}~\bibnamefont{Polchinski}},
  \bibinfo{journal}{JHEP} \textbf{\bibinfo{volume}{1106}}, \bibinfo{pages}{031}
  (\bibinfo{year}{2011}), \eprint{1010.1264}.

\bibitem[{\citenamefont{{Faulkner} et~al.}(2011)\citenamefont{{Faulkner},
  {Liu}, and {Rangamani}}}]{2011JHEP...08..051F}
\bibinfo{author}{\bibfnamefont{T.}~\bibnamefont{{Faulkner}}},
  \bibinfo{author}{\bibfnamefont{H.}~\bibnamefont{{Liu}}}, \bibnamefont{and}
  \bibinfo{author}{\bibfnamefont{M.}~\bibnamefont{{Rangamani}}},
  \bibinfo{journal}{Journal of High Energy Physics}
  \textbf{\bibinfo{volume}{8}}, \bibinfo{eid}{51} (\bibinfo{year}{2011}),
  \eprint{1010.4036}.

\bibitem[{\citenamefont{Lee}(2012{\natexlab{a}})}]{Lee2012}
\bibinfo{author}{\bibfnamefont{S.-S.} \bibnamefont{Lee}},
  \bibinfo{journal}{Nucl.Phys.} \textbf{\bibinfo{volume}{B862}},
  \bibinfo{pages}{781} (\bibinfo{year}{2012}{\natexlab{a}}),
  \eprint{1108.2253}.

\bibitem[{\citenamefont{Lee}(2012{\natexlab{b}})}]{Lee:2012xba}
\bibinfo{author}{\bibfnamefont{S.-S.} \bibnamefont{Lee}},
  \bibinfo{journal}{JHEP} \textbf{\bibinfo{volume}{1210}}, \bibinfo{pages}{160}
  (\bibinfo{year}{2012}{\natexlab{b}}), \eprint{1204.1780}.

\bibitem[{\citenamefont{Lee}(2014)}]{Lee:2013dln}
\bibinfo{author}{\bibfnamefont{S.-S.} \bibnamefont{Lee}},
  \bibinfo{journal}{JHEP} \textbf{\bibinfo{volume}{1401}}, \bibinfo{pages}{076}
  (\bibinfo{year}{2014}), \eprint{1305.3908}.

\bibitem[{\citenamefont{{Chapline}}(2003)}]{2003IJMPA..18.3587C}
\bibinfo{author}{\bibfnamefont{G.}~\bibnamefont{{Chapline}}},
  \bibinfo{journal}{International Journal of Modern Physics A}
  \textbf{\bibinfo{volume}{18}}, \bibinfo{pages}{3587} (\bibinfo{year}{2003}),
  \eprint{gr-qc/0012094}.

\bibitem[{\citenamefont{Lunts et~al.}(2015)\citenamefont{Lunts, Bhattacharjee,
  Miller, Schnetter, Kim, and Lee}}]{Lunts2015}
\bibinfo{author}{\bibfnamefont{P.}~\bibnamefont{Lunts}},
  \bibinfo{author}{\bibfnamefont{S.}~\bibnamefont{Bhattacharjee}},
  \bibinfo{author}{\bibfnamefont{J.}~\bibnamefont{Miller}},
  \bibinfo{author}{\bibfnamefont{E.}~\bibnamefont{Schnetter}},
  \bibinfo{author}{\bibfnamefont{Y.~B.} \bibnamefont{Kim}}, \bibnamefont{and}
  \bibinfo{author}{\bibfnamefont{S.-S.} \bibnamefont{Lee}},
  \bibinfo{journal}{Journal of High Energy Physics}
  \textbf{\bibinfo{volume}{2015}}, \bibinfo{pages}{1} (\bibinfo{year}{2015}),
  ISSN \bibinfo{issn}{1029-8479},
  \urlprefix\url{http://dx.doi.org/10.1007/JHEP08(2015)107}.

\bibitem[{\citenamefont{{Susskind}}(2014)}]{2014arXiv1402.5674S}
\bibinfo{author}{\bibfnamefont{L.}~\bibnamefont{{Susskind}}},
  \bibinfo{journal}{ArXiv e-prints}  (\bibinfo{year}{2014}),
  \eprint{1402.5674}.

\bibitem[{\citenamefont{Heyl et~al.}(2013)\citenamefont{Heyl, Polkovnikov, and
  Kehrein}}]{PhysRevLett.110.135704}
\bibinfo{author}{\bibfnamefont{M.}~\bibnamefont{Heyl}},
  \bibinfo{author}{\bibfnamefont{A.}~\bibnamefont{Polkovnikov}},
  \bibnamefont{and} \bibinfo{author}{\bibfnamefont{S.}~\bibnamefont{Kehrein}},
  \bibinfo{journal}{Phys. Rev. Lett.} \textbf{\bibinfo{volume}{110}},
  \bibinfo{pages}{135704} (\bibinfo{year}{2013}),
  \urlprefix\url{http://link.aps.org/doi/10.1103/PhysRevLett.110.135704}.

\bibitem[{\citenamefont{Heyl}(2015)}]{PhysRevLett.115.140602}
\bibinfo{author}{\bibfnamefont{M.}~\bibnamefont{Heyl}}, \bibinfo{journal}{Phys.
  Rev. Lett.} \textbf{\bibinfo{volume}{115}}, \bibinfo{pages}{140602}
  (\bibinfo{year}{2015}),
  \urlprefix\url{http://link.aps.org/doi/10.1103/PhysRevLett.115.140602}.

\bibitem[{\citenamefont{{Bender}}(2007)}]{2007RPPh...70..947B}
\bibinfo{author}{\bibfnamefont{C.~M.} \bibnamefont{{Bender}}},
  \bibinfo{journal}{Reports on Progress in Physics}
  \textbf{\bibinfo{volume}{70}}, \bibinfo{pages}{947} (\bibinfo{year}{2007}),
  \eprint{hep-th/0703096}.

\bibitem[{\citenamefont{Polchinski}(1984)}]{Polchinski:1983gv}
\bibinfo{author}{\bibfnamefont{J.}~\bibnamefont{Polchinski}},
  \bibinfo{journal}{Nucl.Phys.} \textbf{\bibinfo{volume}{B231}},
  \bibinfo{pages}{269} (\bibinfo{year}{1984}).

\bibitem[{\citenamefont{{Osborn}}(1991)}]{1991NuPhB.363..486O}
\bibinfo{author}{\bibfnamefont{H.}~\bibnamefont{{Osborn}}},
  \bibinfo{journal}{Nuclear Physics B} \textbf{\bibinfo{volume}{363}},
  \bibinfo{pages}{486} (\bibinfo{year}{1991}).

\bibitem[{\citenamefont{{Nakayama}}(2015)}]{2015arXiv150207049N}
\bibinfo{author}{\bibfnamefont{Y.}~\bibnamefont{{Nakayama}}},
  \bibinfo{journal}{ArXiv e-prints}  (\bibinfo{year}{2015}),
  \eprint{1502.07049}.

\bibitem[{\citenamefont{Levin and Nave}(2007)}]{PhysRevLett.99.120601}
\bibinfo{author}{\bibfnamefont{M.}~\bibnamefont{Levin}} \bibnamefont{and}
  \bibinfo{author}{\bibfnamefont{C.~P.} \bibnamefont{Nave}},
  \bibinfo{journal}{Phys. Rev. Lett.} \textbf{\bibinfo{volume}{99}},
  \bibinfo{pages}{120601} (\bibinfo{year}{2007}),
  \urlprefix\url{http://link.aps.org/doi/10.1103/PhysRevLett.99.120601}.

\bibitem[{\citenamefont{{Haegeman} et~al.}(2013)\citenamefont{{Haegeman},
  {Osborne}, {Verschelde}, and {Verstraete}}}]{2013PhRvL.110j0402H}
\bibinfo{author}{\bibfnamefont{J.}~\bibnamefont{{Haegeman}}},
  \bibinfo{author}{\bibfnamefont{T.~J.} \bibnamefont{{Osborne}}},
  \bibinfo{author}{\bibfnamefont{H.}~\bibnamefont{{Verschelde}}},
  \bibnamefont{and}
  \bibinfo{author}{\bibfnamefont{F.}~\bibnamefont{{Verstraete}}},
  \bibinfo{journal}{Physical Review Letters} \textbf{\bibinfo{volume}{110}},
  \bibinfo{eid}{100402} (\bibinfo{year}{2013}), \eprint{1102.5524}.

\bibitem[{\citenamefont{{Evenbly} and {Vidal}}(2015)}]{2015PhRvL.115r0405E}
\bibinfo{author}{\bibfnamefont{G.}~\bibnamefont{{Evenbly}}} \bibnamefont{and}
  \bibinfo{author}{\bibfnamefont{G.}~\bibnamefont{{Vidal}}},
  \bibinfo{journal}{Physical Review Letters} \textbf{\bibinfo{volume}{115}},
  \bibinfo{eid}{180405} (\bibinfo{year}{2015}), \eprint{1412.0732}.

\bibitem[{\citenamefont{{Miyaji} and {Takayanagi}}(2015)}]{2015PTEP.2015g3B03M}
\bibinfo{author}{\bibfnamefont{M.}~\bibnamefont{{Miyaji}}} \bibnamefont{and}
  \bibinfo{author}{\bibfnamefont{T.}~\bibnamefont{{Takayanagi}}},
  \bibinfo{journal}{Progress of Theoretical and Experimental Physics}
  \textbf{\bibinfo{volume}{2015}}, \bibinfo{eid}{073B03}
  (\bibinfo{year}{2015}), \eprint{1503.03542}.

\bibitem[{\citenamefont{Vidal}(2008)}]{PhysRevLett.101.110501}
\bibinfo{author}{\bibfnamefont{G.}~\bibnamefont{Vidal}},
  \bibinfo{journal}{Phys. Rev. Lett.} \textbf{\bibinfo{volume}{101}},
  \bibinfo{pages}{110501} (\bibinfo{year}{2008}),
  \urlprefix\url{http://link.aps.org/doi/10.1103/PhysRevLett.101.110501}.

\bibitem[{\citenamefont{Swingle}(2012)}]{PhysRevD.86.065007}
\bibinfo{author}{\bibfnamefont{B.}~\bibnamefont{Swingle}},
  \bibinfo{journal}{Phys. Rev. D} \textbf{\bibinfo{volume}{86}},
  \bibinfo{pages}{065007} (\bibinfo{year}{2012}),
  \urlprefix\url{http://link.aps.org/doi/10.1103/PhysRevD.86.065007}.

\bibitem[{\citenamefont{{Nozaki} et~al.}(2012)\citenamefont{{Nozaki}, {Ryu},
  and {Takayanagi}}}]{2012JHEP...10..193N}
\bibinfo{author}{\bibfnamefont{M.}~\bibnamefont{{Nozaki}}},
  \bibinfo{author}{\bibfnamefont{S.}~\bibnamefont{{Ryu}}}, \bibnamefont{and}
  \bibinfo{author}{\bibfnamefont{T.}~\bibnamefont{{Takayanagi}}},
  \bibinfo{journal}{Journal of High Energy Physics}
  \textbf{\bibinfo{volume}{10}}, \bibinfo{eid}{193} (\bibinfo{year}{2012}),
  \eprint{1208.3469}.

\bibitem[{\citenamefont{{Qi}}(2013)}]{2013arXiv1309.6282Q}
\bibinfo{author}{\bibfnamefont{X.-L.} \bibnamefont{{Qi}}},
  \bibinfo{journal}{ArXiv e-prints}  (\bibinfo{year}{2013}),
  \eprint{1309.6282}.

\bibitem[{\citenamefont{{Miyaji} et~al.}(2015)\citenamefont{{Miyaji},
  {Numasawa}, {Shiba}, {Takayanagi}, and {Watanabe}}}]{2015PhRvL.115q1602M}
\bibinfo{author}{\bibfnamefont{M.}~\bibnamefont{{Miyaji}}},
  \bibinfo{author}{\bibfnamefont{T.}~\bibnamefont{{Numasawa}}},
  \bibinfo{author}{\bibfnamefont{N.}~\bibnamefont{{Shiba}}},
  \bibinfo{author}{\bibfnamefont{T.}~\bibnamefont{{Takayanagi}}},
  \bibnamefont{and}
  \bibinfo{author}{\bibfnamefont{K.}~\bibnamefont{{Watanabe}}},
  \bibinfo{journal}{Physical Review Letters} \textbf{\bibinfo{volume}{115}},
  \bibinfo{eid}{171602} (\bibinfo{year}{2015}), \eprint{1507.07555}.

\bibitem[{\citenamefont{{Pastawski} et~al.}(2015)\citenamefont{{Pastawski},
  {Yoshida}, {Harlow}, and {Preskill}}}]{2015JHEP...06..149P}
\bibinfo{author}{\bibfnamefont{F.}~\bibnamefont{{Pastawski}}},
  \bibinfo{author}{\bibfnamefont{B.}~\bibnamefont{{Yoshida}}},
  \bibinfo{author}{\bibfnamefont{D.}~\bibnamefont{{Harlow}}}, \bibnamefont{and}
  \bibinfo{author}{\bibfnamefont{J.}~\bibnamefont{{Preskill}}},
  \bibinfo{journal}{Journal of High Energy Physics}
  \textbf{\bibinfo{volume}{6}}, \bibinfo{eid}{149} (\bibinfo{year}{2015}),
  \eprint{1503.06237}.

\bibitem[{\citenamefont{{Marolf}}(2015)}]{2015PhRvL.114c1104M}
\bibinfo{author}{\bibfnamefont{D.}~\bibnamefont{{Marolf}}},
  \bibinfo{journal}{Physical Review Letters} \textbf{\bibinfo{volume}{114}},
  \bibinfo{eid}{031104} (\bibinfo{year}{2015}), \eprint{1409.2509}.

\bibitem[{\citenamefont{{Klebanov} and {Polyakov}}(2002)}]{2002PhLB..550..213K}
\bibinfo{author}{\bibfnamefont{I.~R.} \bibnamefont{{Klebanov}}}
  \bibnamefont{and} \bibinfo{author}{\bibfnamefont{A.~M.}
  \bibnamefont{{Polyakov}}}, \bibinfo{journal}{Physics Letters B}
  \textbf{\bibinfo{volume}{550}}, \bibinfo{pages}{213} (\bibinfo{year}{2002}),
  \eprint{hep-th/0210114}.

\bibitem[{\citenamefont{Das and Jevicki}(2003)}]{Das:2003vw}
\bibinfo{author}{\bibfnamefont{S.~R.} \bibnamefont{Das}} \bibnamefont{and}
  \bibinfo{author}{\bibfnamefont{A.}~\bibnamefont{Jevicki}},
  \bibinfo{journal}{Phys.Rev.} \textbf{\bibinfo{volume}{D68}},
  \bibinfo{pages}{044011} (\bibinfo{year}{2003}), \eprint{hep-th/0304093}.

\bibitem[{\citenamefont{Koch et~al.}(2011)\citenamefont{Koch, Jevicki, Jin, and
  Rodrigues}}]{Koch:2010cy}
\bibinfo{author}{\bibfnamefont{R.~d.~M.} \bibnamefont{Koch}},
  \bibinfo{author}{\bibfnamefont{A.}~\bibnamefont{Jevicki}},
  \bibinfo{author}{\bibfnamefont{K.}~\bibnamefont{Jin}}, \bibnamefont{and}
  \bibinfo{author}{\bibfnamefont{J.~P.} \bibnamefont{Rodrigues}},
  \bibinfo{journal}{Phys.Rev.} \textbf{\bibinfo{volume}{D83}},
  \bibinfo{pages}{025006} (\bibinfo{year}{2011}), \eprint{1008.0633}.

\bibitem[{\citenamefont{Douglas et~al.}(2011)\citenamefont{Douglas, Mazzucato,
  and Razamat}}]{Douglas:2010rc}
\bibinfo{author}{\bibfnamefont{M.~R.} \bibnamefont{Douglas}},
  \bibinfo{author}{\bibfnamefont{L.}~\bibnamefont{Mazzucato}},
  \bibnamefont{and} \bibinfo{author}{\bibfnamefont{S.~S.}
  \bibnamefont{Razamat}}, \bibinfo{journal}{Phys.Rev.}
  \textbf{\bibinfo{volume}{D83}}, \bibinfo{pages}{071701}
  (\bibinfo{year}{2011}), \eprint{1011.4926}.

\bibitem[{\citenamefont{{Pando Zayas} and {Peng}}(2013)}]{2013arXiv1303.6641P}
\bibinfo{author}{\bibfnamefont{L.~A.} \bibnamefont{{Pando Zayas}}}
  \bibnamefont{and} \bibinfo{author}{\bibfnamefont{C.}~\bibnamefont{{Peng}}},
  \bibinfo{journal}{ArXiv e-prints}  (\bibinfo{year}{2013}),
  \eprint{1303.6641}.

\bibitem[{\citenamefont{Leigh et~al.}(2014)\citenamefont{Leigh, Parrikar, and
  Weiss}}]{Leigh:2014tza}
\bibinfo{author}{\bibfnamefont{R.~G.} \bibnamefont{Leigh}},
  \bibinfo{author}{\bibfnamefont{O.}~\bibnamefont{Parrikar}}, \bibnamefont{and}
  \bibinfo{author}{\bibfnamefont{A.~B.} \bibnamefont{Weiss}},
  \bibinfo{journal}{Phys.Rev.} \textbf{\bibinfo{volume}{D89}},
  \bibinfo{pages}{106012} (\bibinfo{year}{2014}), \eprint{1402.1430}.

\bibitem[{\citenamefont{{Leigh} et~al.}(2015)\citenamefont{{Leigh}, {Parrikar},
  and {Weiss}}}]{2015PhRvD..91b6002L}
\bibinfo{author}{\bibfnamefont{R.~G.} \bibnamefont{{Leigh}}},
  \bibinfo{author}{\bibfnamefont{O.}~\bibnamefont{{Parrikar}}},
  \bibnamefont{and} \bibinfo{author}{\bibfnamefont{A.~B.}
  \bibnamefont{{Weiss}}}, \bibinfo{journal}{\prd}
  \textbf{\bibinfo{volume}{91}}, \bibinfo{eid}{026002} (\bibinfo{year}{2015}),
  \eprint{1407.4574}.

\bibitem[{\citenamefont{{Mintun} and {Polchinski}}(2014)}]{2014arXiv1411.3151M}
\bibinfo{author}{\bibfnamefont{E.}~\bibnamefont{{Mintun}}} \bibnamefont{and}
  \bibinfo{author}{\bibfnamefont{J.}~\bibnamefont{{Polchinski}}},
  \bibinfo{journal}{ArXiv e-prints}  (\bibinfo{year}{2014}),
  \eprint{1411.3151}.

\bibitem[{\citenamefont{Vasiliev}(1996)}]{Vasiliev:1995dn}
\bibinfo{author}{\bibfnamefont{M.~A.} \bibnamefont{Vasiliev}},
  \bibinfo{journal}{Int.J.Mod.Phys.} \textbf{\bibinfo{volume}{D5}},
  \bibinfo{pages}{763} (\bibinfo{year}{1996}), \eprint{hep-th/9611024}.

\bibitem[{\citenamefont{Vasiliev}(1999)}]{Vasiliev:1999ba}
\bibinfo{author}{\bibfnamefont{M.~A.} \bibnamefont{Vasiliev}}
  (\bibinfo{year}{1999}), \eprint{hep-th/9910096}.

\bibitem[{\citenamefont{Giombi and Yin}(2010)}]{Giombi:2009wh}
\bibinfo{author}{\bibfnamefont{S.}~\bibnamefont{Giombi}} \bibnamefont{and}
  \bibinfo{author}{\bibfnamefont{X.}~\bibnamefont{Yin}},
  \bibinfo{journal}{JHEP} \textbf{\bibinfo{volume}{1009}}, \bibinfo{pages}{115}
  (\bibinfo{year}{2010}), \eprint{0912.3462}.

\bibitem[{\citenamefont{Vasiliev}(2003)}]{Vasiliev:2003ev}
\bibinfo{author}{\bibfnamefont{M.}~\bibnamefont{Vasiliev}},
  \bibinfo{journal}{Phys.Lett.} \textbf{\bibinfo{volume}{B567}},
  \bibinfo{pages}{139} (\bibinfo{year}{2003}), \eprint{hep-th/0304049}.

\bibitem[{\citenamefont{Maldacena and
  Zhiboedov}(2013{\natexlab{a}})}]{Maldacena:2011jn}
\bibinfo{author}{\bibfnamefont{J.}~\bibnamefont{Maldacena}} \bibnamefont{and}
  \bibinfo{author}{\bibfnamefont{A.}~\bibnamefont{Zhiboedov}},
  \bibinfo{journal}{J.Phys.} \textbf{\bibinfo{volume}{A46}},
  \bibinfo{pages}{214011} (\bibinfo{year}{2013}{\natexlab{a}}),
  \eprint{1112.1016}.

\bibitem[{\citenamefont{Maldacena and
  Zhiboedov}(2013{\natexlab{b}})}]{Maldacena:2012sf}
\bibinfo{author}{\bibfnamefont{J.}~\bibnamefont{Maldacena}} \bibnamefont{and}
  \bibinfo{author}{\bibfnamefont{A.}~\bibnamefont{Zhiboedov}},
  \bibinfo{journal}{Class.Quant.Grav.} \textbf{\bibinfo{volume}{30}},
  \bibinfo{pages}{104003} (\bibinfo{year}{2013}{\natexlab{b}}),
  \eprint{1204.3882}.

\bibitem[{\citenamefont{{Sachs}}(2014)}]{2014PhRvD..90h5003S}
\bibinfo{author}{\bibfnamefont{I.}~\bibnamefont{{Sachs}}},
  \bibinfo{journal}{\prd} \textbf{\bibinfo{volume}{90}}, \bibinfo{eid}{085003}
  (\bibinfo{year}{2014}), \eprint{1306.6654}.

\bibitem[{\citenamefont{{Smerlak}}(2012)}]{2012NJPh...14b3019S}
\bibinfo{author}{\bibfnamefont{M.}~\bibnamefont{{Smerlak}}},
  \bibinfo{journal}{New Journal of Physics} \textbf{\bibinfo{volume}{14}},
  \bibinfo{eid}{023019} (\bibinfo{year}{2012}), \eprint{1104.3303}.

\bibitem[{\citenamefont{Ryu and Takayanagi}(2006)}]{PhysRevLett.96.181602}
\bibinfo{author}{\bibfnamefont{S.}~\bibnamefont{Ryu}} \bibnamefont{and}
  \bibinfo{author}{\bibfnamefont{T.}~\bibnamefont{Takayanagi}},
  \bibinfo{journal}{Phys. Rev. Lett.} \textbf{\bibinfo{volume}{96}},
  \bibinfo{pages}{181602} (\bibinfo{year}{2006}),
  \urlprefix\url{http://link.aps.org/doi/10.1103/PhysRevLett.96.181602}.

\bibitem[{\citenamefont{Hubeny et~al.}(2007)\citenamefont{Hubeny, Rangamani,
  and Takayanagi}}]{1126-6708-2007-07-062}
\bibinfo{author}{\bibfnamefont{V.~E.} \bibnamefont{Hubeny}},
  \bibinfo{author}{\bibfnamefont{M.}~\bibnamefont{Rangamani}},
  \bibnamefont{and}
  \bibinfo{author}{\bibfnamefont{T.}~\bibnamefont{Takayanagi}},
  \bibinfo{journal}{Journal of High Energy Physics}
  \textbf{\bibinfo{volume}{2007}}, \bibinfo{pages}{062} (\bibinfo{year}{2007}),
  \urlprefix\url{http://stacks.iop.org/1126-6708/2007/i=07/a=062}.

\bibitem[{\citenamefont{Van~Raamsdonk}(2010)}]{VanRaamsdonk:2010pw}
\bibinfo{author}{\bibfnamefont{M.}~\bibnamefont{Van~Raamsdonk}},
  \bibinfo{journal}{Gen. Rel. Grav.} \textbf{\bibinfo{volume}{42}},
  \bibinfo{pages}{2323} (\bibinfo{year}{2010}), \bibinfo{note}{[Int. J. Mod.
  Phys.D19,2429(2010)]}, \eprint{1005.3035}.

\bibitem[{\citenamefont{Casini et~al.}(2011)\citenamefont{Casini, Huerta, and
  Myers}}]{Casini2011}
\bibinfo{author}{\bibfnamefont{H.}~\bibnamefont{Casini}},
  \bibinfo{author}{\bibfnamefont{M.}~\bibnamefont{Huerta}}, \bibnamefont{and}
  \bibinfo{author}{\bibfnamefont{R.~C.} \bibnamefont{Myers}},
  \bibinfo{journal}{Journal of High Energy Physics}
  \textbf{\bibinfo{volume}{2011}}, \bibinfo{pages}{1} (\bibinfo{year}{2011}),
  ISSN \bibinfo{issn}{1029-8479},
  \urlprefix\url{http://dx.doi.org/10.1007/JHEP05(2011)036}.

\bibitem[{\citenamefont{Lewkowycz and Maldacena}(2013)}]{Lewkowycz2013}
\bibinfo{author}{\bibfnamefont{A.}~\bibnamefont{Lewkowycz}} \bibnamefont{and}
  \bibinfo{author}{\bibfnamefont{J.}~\bibnamefont{Maldacena}},
  \bibinfo{journal}{Journal of High Energy Physics}
  \textbf{\bibinfo{volume}{2013}}, \bibinfo{pages}{1} (\bibinfo{year}{2013}),
  ISSN \bibinfo{issn}{1029-8479},
  \urlprefix\url{http://dx.doi.org/10.1007/JHEP08(2013)090}.

\bibitem[{\citenamefont{Hu and Lee}(2016)}]{Hu}
\bibinfo{author}{\bibfnamefont{Q.}~\bibnamefont{Hu}} \bibnamefont{and}
  \bibinfo{author}{\bibfnamefont{S.-S.} \bibnamefont{Lee}},
  \bibinfo{journal}{in preparation}  (\bibinfo{year}{2016}).

\bibitem[{\citenamefont{Kachru et~al.}(2008)\citenamefont{Kachru, Liu, and
  Mulligan}}]{PhysRevD.78.106005}
\bibinfo{author}{\bibfnamefont{S.}~\bibnamefont{Kachru}},
  \bibinfo{author}{\bibfnamefont{X.}~\bibnamefont{Liu}}, \bibnamefont{and}
  \bibinfo{author}{\bibfnamefont{M.}~\bibnamefont{Mulligan}},
  \bibinfo{journal}{Phys. Rev. D} \textbf{\bibinfo{volume}{78}},
  \bibinfo{pages}{106005} (\bibinfo{year}{2008}),
  \urlprefix\url{http://link.aps.org/doi/10.1103/PhysRevD.78.106005}.

\bibitem[{\citenamefont{Balasubramanian and
  Narayan}(2010)}]{Balasubramanian2010}
\bibinfo{author}{\bibfnamefont{K.}~\bibnamefont{Balasubramanian}}
  \bibnamefont{and} \bibinfo{author}{\bibfnamefont{K.}~\bibnamefont{Narayan}},
  \bibinfo{journal}{Journal of High Energy Physics}
  \textbf{\bibinfo{volume}{2010}}, \bibinfo{pages}{1} (\bibinfo{year}{2010}),
  ISSN \bibinfo{issn}{1029-8479},
  \urlprefix\url{http://dx.doi.org/10.1007/JHEP08(2010)014}.

\bibitem[{\citenamefont{{Donos} and {Gauntlett}}(2010)}]{2010JHEP...12..002D}
\bibinfo{author}{\bibfnamefont{A.}~\bibnamefont{{Donos}}} \bibnamefont{and}
  \bibinfo{author}{\bibfnamefont{J.~P.} \bibnamefont{{Gauntlett}}},
  \bibinfo{journal}{Journal of High Energy Physics}
  \textbf{\bibinfo{volume}{12}}, \bibinfo{eid}{2} (\bibinfo{year}{2010}),
  \eprint{1008.2062}.

\bibitem[{\citenamefont{{Strominger}}(2001)}]{2001JHEP...10..034S}
\bibinfo{author}{\bibfnamefont{A.}~\bibnamefont{{Strominger}}},
  \bibinfo{journal}{Journal of High Energy Physics}
  \textbf{\bibinfo{volume}{10}}, \bibinfo{eid}{034} (\bibinfo{year}{2001}),
  \eprint{hep-th/0106113}.

\bibitem[{\citenamefont{{Balasubramanian}
  et~al.}(2003)\citenamefont{{Balasubramanian}, {de Boer}, and
  {Minic}}}]{2003AnPhy.303...59B}
\bibinfo{author}{\bibfnamefont{V.}~\bibnamefont{{Balasubramanian}}},
  \bibinfo{author}{\bibfnamefont{J.}~\bibnamefont{{de Boer}}},
  \bibnamefont{and} \bibinfo{author}{\bibfnamefont{D.}~\bibnamefont{{Minic}}},
  \bibinfo{journal}{Annals of Physics} \textbf{\bibinfo{volume}{303}},
  \bibinfo{pages}{59} (\bibinfo{year}{2003}), \eprint{hep-th/0207245}.

\bibitem[{\citenamefont{{Anninos} et~al.}(2011)\citenamefont{{Anninos},
  {Hartman}, and {Strominger}}}]{2011arXiv1108.5735A}
\bibinfo{author}{\bibfnamefont{D.}~\bibnamefont{{Anninos}}},
  \bibinfo{author}{\bibfnamefont{T.}~\bibnamefont{{Hartman}}},
  \bibnamefont{and}
  \bibinfo{author}{\bibfnamefont{A.}~\bibnamefont{{Strominger}}},
  \bibinfo{journal}{ArXiv e-prints}  (\bibinfo{year}{2011}),
  \eprint{1108.5735}.

\end{thebibliography}

\newpage
\begin{appendix}
%%%%%%%%%%%%%%%%%%%%%%%%%%%%%

\section{Entanglement Entropy}

\begin{figure}[h]
\centering
\includegraphics[width = 0.75\columnwidth]{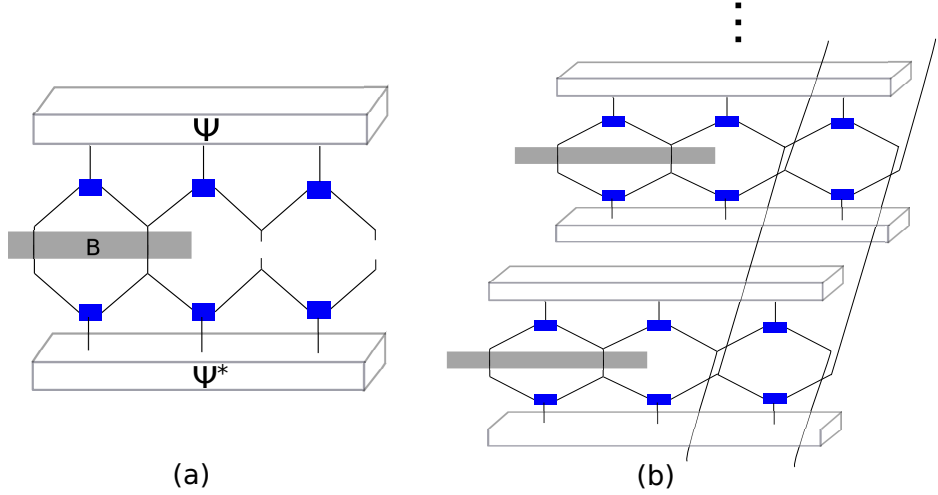}
\caption{
(a) The density matrix $\rho_A$ obtained after tracing out the degrees of freedom in region $B = \bar A$.
(b) A graphical representation of $\tr {\rho_A^n}$.
}
\label{fig:EET}
\end{figure}

In this section, we compute the von Neumann entanglement entropy 
for the state $e^{-z \hat H} \cb t^{(0)}  \rb$ in the large $N$ limit 
to the second order in the hopping field $\bar t_{ij}(z)$.
In the large $N$ limit, we can ignore the fluctuations of $t_{ij}(z)$ in the bulk.
The normalized state can be written as
\bqa
\cb \Psi(z)  \rb & = & \frac{1}{\sqrt{Z_1}}  
 \int D \v{\phi} ~ e^{ ~
 \sum_{ij} \bar t_{ij}(z) \v{\phi}_i^* \cdot \v{\phi}_j 
-\frac{\lambda}{N}\sum_i\left(\v{\phi}^*_i\cdot\v{\phi}_i\right)^2
} \cb \v{\phi} \rb,
\eqa
where $Z_1$ is the normalization factor,
\bqa
Z_1 & = & \int D \v{\phi} ~ e^{ ~
 \sum_{ij} [ \bar t_{ij}(z) + \bar t_{ji}^*(z) ] \v{\phi}_i^* \cdot \v{\phi}_j 
-\frac{2 \lambda}{N}\sum_i\left(\v{\phi}^*_i\cdot\v{\phi}_i\right)^2
}.
\eqa
The density matrix $\rho_A$ of region $A$ is obtained by integrating out $\v{\phi}_i$ in its complement $B \equiv \bar A$ (\fig{fig:EET}(a)).
The von Neumann entanglement entropy is defined to be 
\bqa
S_E = - \lim_{n \rightarrow 1} \frac{ \tr{ \rho_A^n } - 1 }{ n-1},
\label{eq:S_E}
\eqa
where $\tr{ \rho_A^n }$ is a partition function 
for $n$ copies of the system with twisted boundary conditions
as is represented in \fig{fig:EET}(b).
To the second order in $\bar t_{ij}$, we obtain
\bqa
\tr{ \rho_A^n } = \frac{
1 + \frac{ N n}{2} \sum_{i \neq j} 
( \bar t_{ij} + \bar t_{ji}^* )  
( \bar t_{ji} + \bar t_{ij}^* )  
G_0^2 
+ N \sum^{'}_{i, j} \left[ |\bar t_{ij}|^{2n} G_0^{n} - n |\bar t_{ij}|^2 G_0  \right]  + O(\bar t^3)}{
1 + \frac{ N n}{2} \sum_{i \neq  j}^{'} 
( \bar t_{ij} + \bar t_{ji}^* )  
( \bar t_{ji} + \bar t_{ij}^* )  
G_0^2 +O(\bar t^3) }.
\label{eq:rhon}
\eqa
Here all $\bar t_{ij}$'s represent the saddle point values at $z$.
$\sum_{i \neq j}$ sums over ordered pairs of all distinct sites
whereas $\sum^{'}_{i,j}$ sums over ordered pairs 
of which one site belongs to $A$ and the other to $B$.
$G_0$ is the on-site propagator given by
\bqa
G_0 = \frac{1}{N} 
\frac{
\int d\v{\phi}  ~   ( \v{\phi}^* \cdot \v{\phi})  ~
e^{ 2 \bar t_0(z) \v{\phi}^*\cdot\v{\phi}  -  \frac{2 \lambda}{N} (\v{\phi}^*\cdot\v{\phi})^2}  }{\int d\v{\phi}  ~ e^{ 2 \bar t_0(z) \v{\phi}^*\cdot\v{\phi}  -  \frac{2 \lambda}{N} (\v{\phi}^*\cdot\v{\phi})^2}      }
\label{eq:G0}
\eqa
with $\bar t_0(z) = \frac{ \bar t_{ii}(z) + \bar t_{ii}^*(z) }{2}$.
\eq{eq:S_E} and \eq{eq:rhon} lead to \eq{eq:EE}.

\end{appendix}

\end{document}